\documentclass[fleqn,usenatbib]{mnras}
\usepackage{newtxtext,newtxmath}
\usepackage{lipsum}

\DeclareRobustCommand{\VAN}[3]{#2}
\let\VANthebibliography\thebibliography
\def\thebibliography{\DeclareRobustCommand{\VAN}[3]{##3}\VANthebibliography}
\usepackage[T1]{fontenc}

\usepackage{graphicx}	
\usepackage{amsmath}	

\def\lsim{\hbox{\rlap{\raise 0.425ex\hbox{$<$}}\lower 0.65ex\hbox{$\sim$}}}
\def\gsim{\hbox{\rlap{\raise 0.425ex\hbox{$>$}}\lower 0.65ex\hbox{$\sim$}}}
\def\arcmin{\hbox{$^\prime$}}
\def\arcsec{\hbox{$^{\prime\prime}$}}

\defcitealias{siebert19}{S19}
\defcitealias{fk11}{FK11}
\defcitealias{fsk11}{FSK11}
\defcitealias{Fitzpatrick}{Fitzpatrick}

\title[The Type Icn SN~2022ann]{SN~2022ann: A type Icn supernova from a dwarf galaxy that reveals helium in its circumstellar environment}

\def\ucsc{1}
\def\psu{2}
\def\dark{3}
\def\uiuc{4}
\def\ifa{5}
\def\gemini{6}
\def\umn{7}
\def\ciera{8}
\def\ncu{9}
\def\lco{10}
\def\ucsb{11}
\def\jhu{12}
\def\stsci{13}
\def\het{14}
\def\mel{15}
\def\arc{16}
\def\cbpf{17}
\def\rut{18}
\def\ucb{19}
\def\ice{20}
\def\ieec{21}
\def\sp{22}
\def\seti{23}
\def\ufrgs{24}
\def\noao{25}
\def\qub{26}

\author[K.~W.~Davis et al.]{K.~W.~Davis,$^{\ucsc}$\thanks{E-mail: kywdavis@ucsc.edu}
K.~Taggart,$^{\ucsc}$ %
S.~Tinyanont,$^{\ucsc}$
R.~J.~Foley,$^{\ucsc}$
V.~A.~Villar,$^{\psu}$
L.~Izzo,$^{\dark}$
C.~R.~Angus,$^{\dark}$
\newauthor
M.~J.~Bustamante-Rosell,$^{\ucsc}$
D.~A.~Coulter,$^{\ucsc}$
N.~Earl,$^{\uiuc}$
D.~Farias,$^{\dark}$
J.~Hjorth,$^{\dark}$
M.~E.~Huber,$^{\ifa}$
\newauthor
D.~O.~Jones,$^{\ucsc,\gemini}$\thanks{Einstein Fellow}
P.~L.~Kelly,$^{\umn}$
C.~D.~Kilpatrick,$^{\ciera}$
D.~Langeroodi,$^{\dark}$
H.-Y.~Miao,$^{\ncu}$
C.~M.~Pellegrino,$^{\lco,\ucsb}$
\newauthor
E.~Ramirez-Ruiz,$^{\ucsc}$
C.~L.~Ransome,$^{\psu}$
S.~Rest,$^{\jhu}$
S.~N.~Sharief,$^{\uiuc}$
M.~R.~Siebert,$^{\stsci}$
G.~Terreran,$^{\lco}$
\newauthor
I.~M.~Thornton,$^{\psu}$
G.~R.~Zeimann,$^{\het}$
K.~Auchettl,$^{\ucsc,\mel,\arc}$
C.~R.~Bom,$^{\cbpf}$
T~.B.~Brink,$^{\ucb}$
J.~Burke,$^{\lco,\ucsb}$
\newauthor
Y.~Camacho-Neves,$^{\rut}$
K.~C.~Chambers,$^{\ifa}$
T.~J.~L.~de~Boer,$^{\ifa}$
L.~DeMarchi,$^{\ciera}$
A.~V.~Filippenko,$^{\ucb}$
\newauthor
L.~Galbany,$^{\ice,\ieec}$
C.~Gall,$^{\dark}$
H.~Gao,$^{\ifa}$
F.~R.~Herpich,$^{\sp}$
D.~A.~Howell,$^{\lco,\ucsb}$
W.~V.~Jacobson-Galan,$^{\ucb}$
\newauthor
S.~W.~Jha,$^{\rut}$
A.~Kanaan,$^{10}$
N.~Khetan,$^{\dark}$
L.~A.~Kwok,$^{\rut}$
Z.~Lai,$^{\ucsc}$
C.~Larison,$^{\rut}$
C.-C.~Lin,$^{\ifa}$
K.~C.~Loertscher,$^{\ucsc}$
\newauthor
E.~A.~Magnier,$^{\ifa}$
C,~McCully,$^{\lco}$
P.~McGill,$^{\ucsc}$
M.~Newsome,$^{\lco,\ucsb}$
E.~Padilla~Gonzalez,$^{\lco,\ucsb}$
Y.-C.~Pan,$^{\ncu}$
\newauthor
A.~Rest,$^{\stsci,\jhu}$
J.~Rho,$^{\seti}$
T.~Ribeiro,$^{\ufrgs}$
A.~Santos,$^{\cbpf}$
W.~Schoenell,$^{\noao}$
S.~N.~Sharief,$^{\uiuc}$
K.~W.~Smith,$^{\qub}$
\newauthor
R.~J. Wainscoat,$^{\ifa}$
Q.~Wang,$^{\jhu}$
S.~K.~Yadavalli,$^{\psu}$
Y.~Zenati,$^{\jhu}$\thanks{ISEF Postdoctoral Fellow}
W.~Zheng$^{\ucb}$\\
Affiliations are listed in Appendix \ref{appendix:affil}.
}
\date{Accepted XXX. Received YYY; in original form ZZZ}

\pubyear{2022}

\begin{document}
\label{firstpage}
\pagerange{\pageref{firstpage}--\pageref{lastpage}}
\maketitle

\begin{abstract}
  We present optical and near-infrared (NIR) observations of the Type Icn supernova (SN~Icn) 2022ann, the fifth member of its newly identified class of SNe. Its early optical spectra are dominated by narrow carbon and oxygen P-Cygni features with absorption velocities of $\sim$800~km~s$^{-1}$; slower than other SNe~Icn and indicative of interaction with a dense, H/He-poor circumstellar medium (CSM) that is outflowing slower than a typical Wolf-Rayet wind velocity of $>$1000~km~s$^{-1}$. We identify helium in NIR spectra obtained two weeks after maximum and in optical spectra at three weeks, demonstrating that the CSM is not fully devoid of helium. We never detect broad spectral features from SN ejecta, including in spectra extending to the nebular phase, a unique characteristic among SNe~Icn. Compared to other SNe~Icn, SN~2022ann has a low luminosity, with a peak $o$-band absolute magnitude of $\sim -17.7$, and evolves slowly. We model the bolometric light curve and find it is well-described by $\sim$ 1.7~M$_{\odot}$ of SN ejecta interacting with 0.2~M$_{\odot}$ of CSM. We place an upper limit of 0.04~M$_{\odot}$ of $^{56}$Ni synthesized in the explosion. The host galaxy is a dwarf galaxy with a stellar mass of 10$^{7.34}$ M$_\odot$ (implied metallicity of log$(Z/{\rm Z}_\odot) \approx 0.10$) and integrated star-formation rate of $\log({\rm SFR}) = -2.20$~M$_{\odot}$~yr$^{-1}$; both lower than 97\% of the galaxies observed to produce core-collapse supernovae, although consistent with star-forming galaxies on the galaxy Main Sequence. The low CSM velocity, nickel and ejecta masses, and likely low-metallicity environment disfavour a single Wolf-Rayet progenitor star. Instead, a binary companion star is likely required to adequately strip the progenitor before explosion and produce a low-velocity outflow. The low CSM velocity may be indicative of the outer Lagrangian points in the stellar binary progenitor, rather than from the escape velocity of a single Wolf-Rayet-like massive star.

\end{abstract}

\begin{keywords}
transients: supernovae, binaries, stars: massive
\end{keywords}

\section{Introduction}\label{sec:intro}

Massive stars $\gtrsim$8~M$_{\odot}$ typically end their lives in terminal explosions known as core-collapse supernovae (CCSNe). Some massive stars, as a result of either strong stellar winds or interaction with a companion, are stripped of their hydrogen envelopes \citep{Woosley1995, Eldridge2008, Tauris2013, Tauris2015}, producing a stripped-envelope SN (SESN) of Type Ib. Further stripping can remove the helium envelope, exposing the remaining carbon/oxygen core. If such a star exploded, the resulting SN would lack signatures of hydrogen and helium, producing a Type Ic SN (SN~Ic; for a review of spectroscopic classification, see \citealt{Filippenko1997}). However, carbon burning lasts only $\sim$100~yr for a star with a zero-age main sequence mass of $\sim$25~M$_{\sun}$. If we assume that the lack of observed helium is due to the absence of helium itself, this sets a stringent timescale of no more than decades (or perhaps a few centuries) before the explosion for the entirety of the stripping to occur. Another possible way to produce SNe Ic may be to ``hide'' the helium. The excitation of helium requires high-energy photons, such as gamma-rays produced by nearby radioactive iron-group elements \citep{Filippenko1995, Clocchiatti1996, Dessart2012, Hachinger2012, Teffs2020, Williamson2021}. If the SN produces a small amount of iron-group elements, or if the helium and iron-group elements are physically separated from one another in the ejecta, it is possible for the helium to remain ``hidden''. 

A fraction of CCSNe display relatively narrow emission features in their optical spectra indicating that the SN is interacting with a dense, close-in circumstellar medium (CSM).  While passing through the CSM, the SN shock will produce X-rays that travel in front of the shock, exciting the unshocked CSM.  This material will emit through recombination lines and the spectral features will have a velocity width corresponding to the outflow velocity of the CSM, which is orders of magnitude lower than the velocity of the SN ejecta.  Starting with the identification of SNe with narrow hydrogen lines, the classes of these objects are denoted with an ``n'' \citep[i.e., SN~IIn for the hydrogen case;][]{Filippenko89, Schlegel1990}.  SNe with narrow He and weak or absent H lines were discovered over two decades ago \citep{Matheson00}; however, the discovery of SN~2006jc with its narrow He lines \citep[e.g.,][]{Foley07}, luminous outburst 2~yr before explosion \citep[e.g.,][]{Pastorello07}, and fast dust formation \citep{Smith07} was the first ``SN~Ibn.''

The recently discovered Type~Icn class \citep{Fraser2021, Gal-Yam2022, Perley2022, Pellegrino2022}, which have strong, narrow O and C lines but weak or absent H and He lines, presents additional complications to the stripping mechanism. Similarly to SNe~Ibn, SNe~Icn have narrow emission features indicative of circumstellar interaction (CSI). However, their spectra show emission that is primarily (and in some cases, exclusively) from carbon and oxygen at early times, with little to no indication of hydrogen (unlike SNe~IIn) or helium (unlike SNe~Ibn).  The lack of detected helium in the CSM surrounding an SN~Icn is particularly confounding since it would require the removal of He from the surface of the star significantly before explosion such that it is no longer present in the CSM.

After the discovery of the first SN~Icn, SN~2019hgp \citep{Gal-Yam2022}, the community has discovered one additional member of the subclass \citep[SN~2021csp;][]{Perley2022, Fraser2021} and reclassified two older SNe as SN~Icn \citep[SNe~2019jc~and~2021ckj;][]{Pellegrino2022}, giving a total of four confirmed SNe~Icn. All SNe~Icn with early-time spectroscopy show similar P-Cygni profiles from carbon and oxygen lines with absorption velocities ($\sim$1000--2000~km~s$^{-1}$), consistent with that of SNe~Ibn and Wolf-Rayet (WR) winds. The qualitative similarities between SNe~Ibn and Icn are also analogous to the differences between WR subtypes, namely WN (He-rich/N-rich) and WO (C-rich/He-poor) stars, respectively. Because of this, WR stars are commonly invoked as progenitors for SESNe, including SNe~Ibn~and~ Icn. WR winds, in combination with a stage of enhanced mass-loss shortly before explosion, can explain the observed properties of SNe~Ibn~and~Icn \citep[e.g.,][]{Gal-Yam2022}.

However, several studies have suggested that WR stars cannot be the sole progenitor channel for SESNe \citep[e.g.,][]{Eldridge2013}.  In particular, the local stellar environments of some SESNe appear to be older than one would expect for a WR star \citep{Anderson2012, Sun2022}. Within the SN~Icn subclass, SN~2019jc exploded in the outskirts of its host galaxy in a region with  low star-formation-rate (SFR) density \citep{Pellegrino2022}. Moreover, the progenitor stars of several SNe~IIb \citep[SESNe with a low-mass H envelope;][]{Filippenko88, Filippenko93} show evidence of being in binary systems \citep[e.g.,][]{Aldering1994, Crockett2008, VanDyk2011, Bersten2012, Kilpatrick2017}, with one having a confirmed binary companion \citep{Maund2004, Fox2014}.  Both of the SNe~Ib having pre-explosion detections of their progenitor systems (iPTF13bvn and SN~2019yvr) are consistent with binary progenitor systems \citep[although see \citealt{Groh2013}]{Cao2013, Eldridge2016, Folatelli2016, Kilpatrick2021}.  SN~2017ein, the first SN~Ic with a possible pre-explosion detection of its progenitor system, has a luminous progenitor system inconsistent with a single WR star, but consistent with a WR star in a binary with a luminous B-type star \citep{Kilpatrick2018, VanDyk2018}. Additionally, there is a late-time source spatially consistent with SN~2006jc that could be a binary companion \citep{Maund2016}.  It remains unclear what fraction of SESNe arise from single WR stars or from binary systems.

In this work, we present observations and analysis of the fifth member of the SN~Icn class, SN~2022ann.  It was discovered in SDSS~J101729.72--022535.6 by ATLAS \citep{Tonry2022} on 2022 January 27.497 (UT dates are used throughout this paper) and classified by us as an SN~Icn on 2022 February 6.283 \citep{Davis2022}. Detailed study of the well-observed SN~2022ann presents an opportunity to examine the progenitor of a rare but important CCSN with extreme mass loss prior to explosion.  This fifth example of an SN~Icn also allows for a study of this small but growing and diverse SN class.  Here we present optical photometry and optical/near-infrared (NIR) spectroscopy of SN~2022ann, as well as a study of its host galaxy.

The paper is organised as follows. Section~\ref{sec:observations} presents the discovery of SN~2022ann, our observational follow-up campaign, and the resulting optical and NIR spectroscopy and optical imaging. In Section~\ref{sec:results}, we provide further insights on the data, model the light curve with CSI and $^{56}$Ni-decay models, and conduct a study of the host galaxy. In Section~\ref{sec:disc}, we discuss the presence of helium in SNe~Icn, the growing heterogeneity observed in the class of SNe~Icn, and the progenitor system for SN~2022ann. We present our conclusions in Section~\ref{sec:conc}.

\section{Observations} \label{sec:observations}

\begin{figure*}
\includegraphics[width=0.98\textwidth]{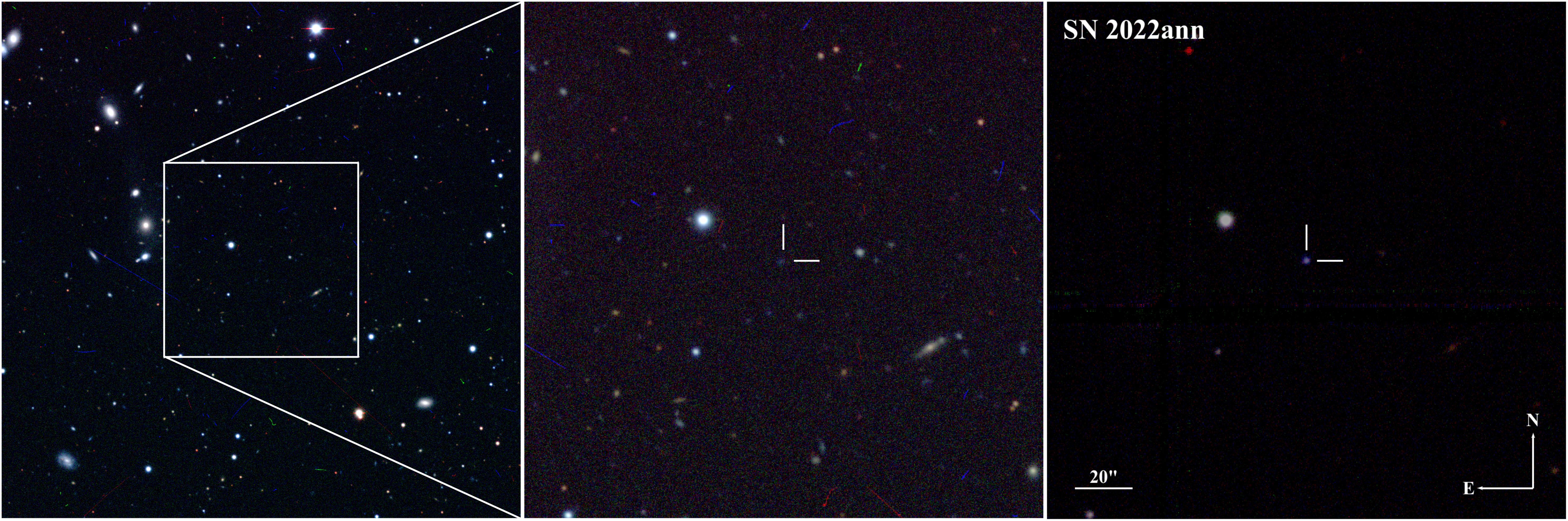}
\caption{Finder charts of SN~2022ann (right) and its host galaxy, SDSS~J101729.72--022535.6 (centre and left).  North is up and east is to the left.   The left panel is an $8\arcmin\ \times 8\arcmin$ finder chart centred on the position of SN~2022ann from archival $grz$ DECam images.  To the east are members of the V1CG 662 galaxy group at $z = 0.0495$ \citep{Lee2017}.  The DECam images were processed by the Dark Energy Spectroscopic Instrument Legacy Imaging Surveys \citep{Dey2019}.  The centre panel is a $3\arcmin\ \times 3\arcmin$ close-up view with the location of SN~2022ann marked.  SN~2022ann is offset slightly to the northwest of SDSS~J101729.72--022535.6.  The right panel is created from  $3\arcmin\ \times 3\arcmin$ Pan-STARRS $gri$ images from 3 Feb.,  30 Jan., and 3 Feb. that are mapped to the blue, red, and green channels of the image,  respectively.\label{fig:finder}}
\end{figure*}

SN~2022ann is offset from the nucleus of its host galaxy, SDSS~J101729.72--022535.6, by 0.96\farcs (see Figure \ref{fig:finder}). From host-galaxy H$\alpha$ emission in our latest spectra of SN~2022ann, we measure a redshift $z = 0.04938$. We discuss the host galaxy in detail in Section~\ref{sec:host}.

Assuming a standard $\Lambda$CDM cosmology (H$_{0} = 70$ km s$^{-1}$ Mpc$^{-1}$, $\Omega_{M} = 0.3$, $\Omega_{\Lambda} = 0.7$), the host-galaxy redshift corresponds to a distance of 218~Mpc, which we adopt as the distance to SN~2022ann throughout this paper.  We also adopt a foreground reddening of $E(B-V)_{\rm MW} = 0.034$~mag \citep{Schlafly2011}. Because of the lack of narrow absorption lines from the interstellar medium in spectra of SN~2022ann, we assume the host galaxy extinction to be negligible. While we lack the necessary data to support this claim further (i.e. by measuring a Balmer decrement), the similar colours to other SNe~Icn (see Section \ref{sec:photanalysis}) suggest that SN~2022ann is not significantly extinguished. Basic parameters for SN~2022ann and its host galaxy are shown in Table~\ref{tbl:paramtable}. Next, we describe our observations and data reduction.

\begin{table}
    \centering
    \caption{Basic observational parameters of SN~2022ann. Presented apparent magnitudes are not extinction-corrected. The $o$ band is a wide-pass filter that covers roughly 5600--8200~\AA, comparable to the combined wavelength coverage of the $r$ and $i$ filters. There is no indication of host-galaxy reddening at the SN location.} 
     \begin{tabular}{lr}
\hline
\hline
Time of First Detection (MJD) & 59604.5\\
Estimated Time of Explosion (MJD) & 59600.5\\
Estimated Time of Maximum (MJD) & 59613.5\\
$\textrm{RA (J2000)}$ & $10^{\textrm{h}}17^{\textrm{m}}29.66^{\textrm{s}}$\\
$\textrm{Dec (J2000)}$ & $-02^{\circ}25'35\farcs45$\\
Redshift & $0.04938 \pm 0.0004$\\  
$E(B-V)_{\textrm{MW}}$ & 0.034 $\pm$ 0.001~mag\\
$E(B-V)_{\textrm{host}}$ & 0\\
$m_{o}^{\mathrm{discovery}}$ & 19.22 $\pm$ 0.11~mag\\
$M_{o}^{\mathrm{discovery}}$ & $-17.47$~mag\\
$m_{o}^{\mathrm{peak}}$ & 19.00 $\pm$ 0.069~mag\\
$M_{o}^{\mathrm{peak}}$ & $-17.69$~mag \\
\hline
\end{tabular}\label{tbl:paramtable}
\end{table}

\subsection{Photometry}\label{sec:photobs}

\begin{figure*}
\centering
\includegraphics[width=\textwidth]{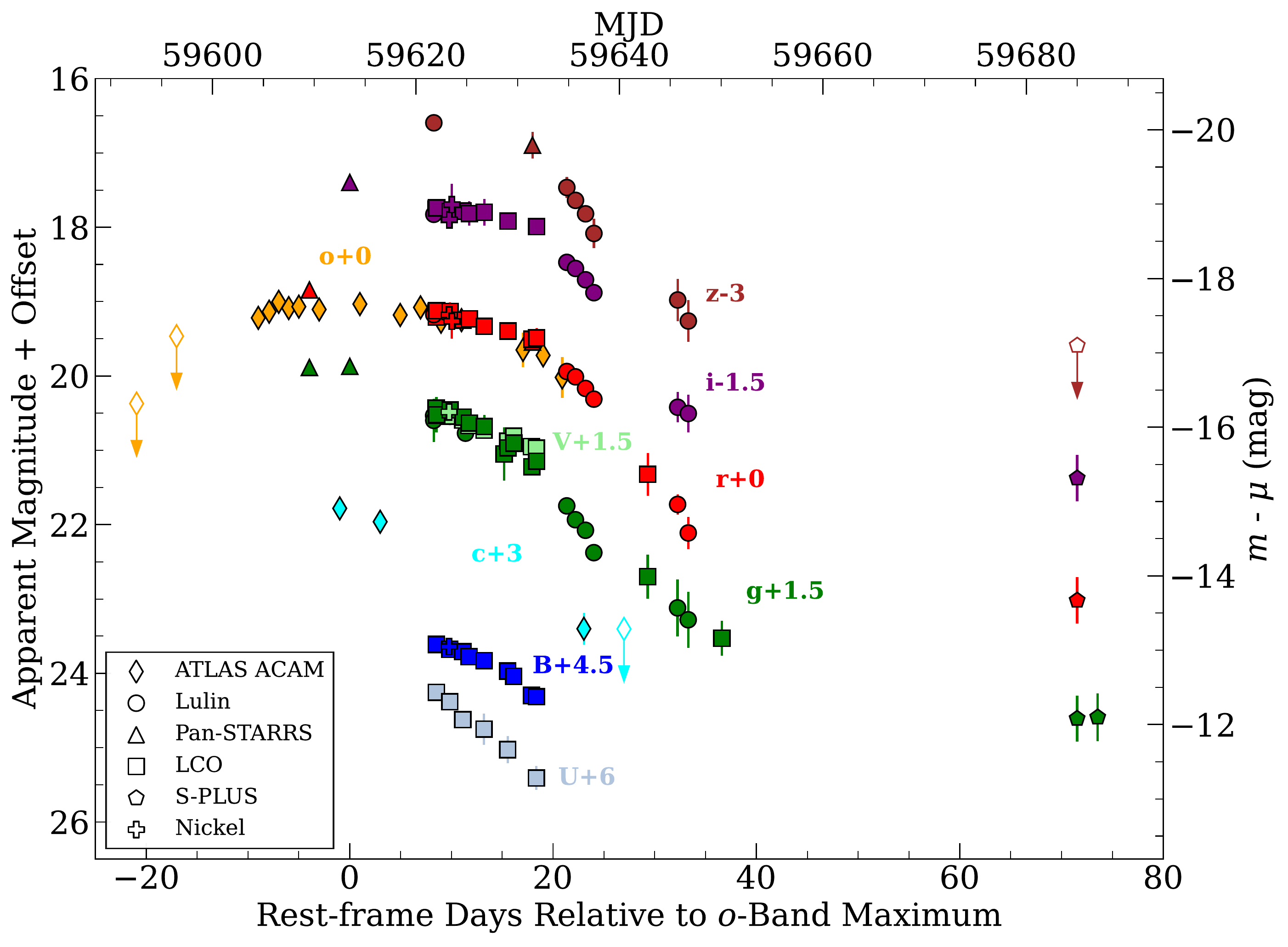}
\caption{Optical light curves of SN~2022ann from ATLAS (diamonds), Lulin (circles), Pan-STARRS (triangles), LCO (squares), S-PLUS (pentagons), and Lick/Nickel (plus-signs). The phase is given relative to the $o$-band maximum. $U\!BV$ magnitudes are reported in the Vega system \citep{Johnson53}, while all others are in AB \citep{Oke83}. 3$\sigma$ nondetections are shown as white-filled points with a downward arrow.\label{fig:photometry}}
\end{figure*}

We present our photometric observations of SN~2022ann in Figure~\ref{fig:photometry} and Table~\ref{tab:phottable}, where our data continue for 90~days after discovery. We define the time of explosion, $t_{0}$, as the midpoint between the last pre-explosion nondetection and the first detection.  For SN~2022ann, there are two relevant ATLAS nondetections with which we can estimate the time of explosion: a deeper, but earlier nondetection on 2022 January 13.5 (MJD 59592.5) and a shallower, but later nondetection on 2022 January 17.5 (MJD 59596.5) with corresponding limiting magnitudes of $o = 20.37$ and 19.47~mag, respectively. In our analysis, we use the later nondetection to estimate the time of explosion as 2022 January 21.5 (MJD 59600.5).  Doing so results in a rise time in the $o$-band of $\sim$10~days, which is similar to that of other SNe~Ibn/Icn, but we note that the nondetections are not particularly constraining.  As we do not perform any detailed modeling that relies on a precise measurement of $t_{0}$, the current constraint is sufficient for our analysis presented here. We estimate the time of maximum brightness using the $o$-band light curve owing to its coverage at early times. The light curve is very flat around peak (see Section~\ref{sec:results}). We estimate the time of peak brightness to be the midpoint of this flat region, which yields a $t_{\rm peak}$ of 2022 February 03.5 (MJD 59613.5). Phases for SN~2022ann in this paper are given relative to the $o$-band time of maximum unless stated otherwise.

SN~2022ann was observed in the $c$ and $o$ bands by ATLAS between $-8$ and 21~days. We use the ATLAS forced photometry server \citep{Tonry2018, Smith2020, Shingles21} to recover the difference-image photometry for SN~2022ann. To remove erroneous measurements and have significant SN flux detection at the location of SN~2022ann, we apply several cuts on the total number of individual data points and nightly averaged data. Our first cut uses the $\chi^{2}$ and uncertainty values of the point-spread-function (PSF) fitting to remove discrepant data. We then obtain forced photometry of eight control light curves located in a circular pattern around the location of the SN with a radius of 17\arcsec. The flux of these control light curves is expected to be consistent with zero within the uncertainties, and any deviation from zero would indicate that there are either unaccounted systematic biases or underestimated uncertainties. We search for such deviations by calculating the weighted mean of the set of control light-curve measurements for a given epoch after removing any $>3\sigma$ outliers (for a more detailed discussion, see Rest et~al., in prep.\footnote{\url{https://github.com/srest2021/atlaslc}}). If the weighted mean of these photometric measurements is inconsistent with zero, we flag and remove those epochs from the SN light curve. This method allows us to identify potentially incorrect measurements without using the SN light curve itself. We then bin the SN~2022ann light curve by calculating a 3$\sigma$-cut weighted mean for each night (ATLAS typically has four epochs per night), excluding the flagged measurements from the previous step.

We also observed SN~2022ann with the Lulin Compact Imager on the 1~m telescope at Lulin Observatory from 2022 February 11 to 2022 March 8 in the $griz$ bands.  The images were calibrated using bias and flat-field frames following standard procedures.  To account for background emission due to its host galaxy, we subtracted Pan-STARRS 3$\pi$ cutout images \citep{Magnier16} using the image convolution and subtraction software, {\tt HOTPANTS} \citep{Becker15}.  We then performed forced photometry in each frame using {\tt DoPhot} \citep{Schechter93} within {\tt photpipe} \citep{Rest05}.  The photometry was calibrated using Pan-STARRS $griz$ local standard stars \citep{Flewelling16}.

We also obtained $uBgVri$-band images of SN~2022ann using Sinistro cameras on Las Cumbres Observatory (LCO) 1.0~m telescopes from 2022 February 11 to 2022 March 8. The photometry was collected as part of the Global Supernova Project (GSP) and from separate programs (PIs R. Foley and C. Kilpatrick). GSP data were reduced using the \texttt{lcogtsnpipe} pipeline \citep{Valenti2016}, which extracts PSF magnitudes after calculating zero-points and colour terms \citep{Stetson1987}. $U\!BV$ photometry was calibrated to Vega magnitudes using \citet{Landolt1992} standard fields, and $gri$ photometry was calibrated to AB magnitudes \citep{Smith2002} using Sloan Digital Sky Survey (SDSS) catalogues. Background subtraction was performed using {\tt HOTPANTS} with template images obtained after the SN had faded. Data from separate programs were reduced using the same procedures described above for the Lulin telescope, but with SkyMapper images as templates and SkyMapper local standard stars \citep{Wolf18} in the $u$ band. Both reduction methods produce results consistent with one another when comparing epochs with overlapping coverage.

Imaging was also obtained on 2022 February 13 in the $BVri$ bands with the 1~m Nickel telescope at Lick Observatory. The images were calibrated using bias and sky flat-field frames following standard procedures and were subtracted with a reference image obtained on 2022 March 12. PSF photometry was performed, and photometry was calibrated relative to Pan-STARRS photometric standards \citep{Flewelling16}.

We also observed SN~2022ann with the T80 telescope via the Southern Photometric Local Universe Survey \citep[S-PLUS;][]{deOliveira19} Transient Extension Program \citep[STEP;][]{2022TNSAN.178}.  We used the standard S-PLUS observation strategy described by \citet{deOliveira19} to observe SN~2022ann in $griz$ from 2022 April 16 to 2022 May 5.  These data were initially processed for pixel-level corrections using the {\tt JYPE} pipeline \citep{Cristobal-Hornillos14}.  We then reduced all STEP data using {\tt photpipe} \citep{Rest05}, including masking, regridding each image to a common image centre in {\tt SWarp} \citep{swarp}, PSF photometry with {\tt DoPhot} \citep{Schechter93}, and photometric calibration using Pan-STARRS DR2 standard stars observed in the same field as SN~2022ann \citep{Flewelling16}.  Next, we subtracted $griz$ template images obtained from the same telescope and reduced in the same way from 2022 June 15 using {\tt HOTPANTS}.  Finally, we performed forced photometry in the subtracted images at the site of SN\,2022ann using a custom version of {\tt DoPhot}.

Additional $griz$-band imaging was obtained through the Young Supernova Experiment (YSE) \citep{Jones2021} with the Pan-STARRS telescope \citep[PS1;][]{Kaiser2002} between 2022 January 30 and 2022 February 21. The YSE photometric pipeline is based on {\tt photpipe} \citep{Rest05}. Each image template was taken from stacked PS1 exposures, with most of the input data from the PS1 3$\pi$ survey. All images and templates are resampled and astrometrically aligned to match a skycell in the PS1 sky tessellation. An image zero-point is determined by comparing PSF photometry of the stars to updated stellar catalogues of PS1 observations \citep{Chambers17}. The PS1 templates are convolved with a three-Gaussian kernel to match the PSF of the nightly images, and the convolved templates are subtracted from the nightly images with {\tt HOTPANTS} \citep{Becker15}. Finally, a flux-weighted centroid is found for each SN position and PSF photometry is performed using forced photometry. The nightly zero-point is applied to the photometry to determine the brightness of the SN for that epoch.

SN~2022ann was also observed with the Ultraviolet Optical Telescope (UVOT; \citealt{Roming05}) onboard the Neil Gehrels \emph{Swift} Observatory \citep{Gehrels04} on 2022 July 08.95 (MJD 59768.95, 155.5 days after $o$-band maximum). We performed aperture photometry with a 5$\arcsec$ region with \texttt{uvotsource} within HEAsoft v6.26\footnote{We used the calibration database (CALDB) version 20201008.}, following the standard guidelines from \cite{Brown14}. We detect no UV emission from the SN in this image and obtain a 3$\sigma$ limiting magnitude of 23.016 in the $w2$-band. This nondetection is not shown in Figure~\ref{fig:photometry}, but is made available in Table~\ref{tab:phottable}.

\subsection{Spectroscopy}\label{sec:specobs}

We spectroscopically followed SN~2022ann starting at 2.8~days and continuing through 80.8~days after maximum light.  The optical spectra were obtained with the Kast dual-beam spectrograph \citep{KAST} on the Lick Shane 3~m telescope, the Goodman spectrograph \citep{SOAR} on the NOIRLab 4.1~m Southern Astrophysical Research (SOAR) telescope at Cerro Pach\'on, the Alhambra Faint Object Spectrograph (ALFOSC) on the Nordic Optical Telescope (NOT), second-generation Low Resolution Spectrograph \citep[LRS2;][]{LRS2} on the Hobby-Eberly Telescope (HET), Binospec on the MMT, the Gemini Multi-Object Spectrograph \citep[GMOS;][]{GMOS} on the 8.1~m Gemini-South telescope, and the Low-Resolution Imaging Spectrograph \citep[LRIS;][]{LRIS} on the 10~m Keck~I telescope. A log of all observations is presented in Table~\ref{tab:speclog}. Observations were taken at low airmass (< 2) with the slit at the parallactic angle \citep{Filippenko82} (unless the instrument was equipped with an Atmospheric-Dispersion Corrector, in which case the slit was aligned through SN~2022ann and the nucleus of its host galaxy).  

To reduce the Kast, Goodman, GMOS, and LRIS spectral data, we used the {\tt UCSC Spectral Pipeline}\footnote{\url{https://github.com/msiebert1/UCSC\_spectral\_pipeline}} \citep{siebert19}, a custom data-reduction pipeline based on procedures outlined by \citet{Foley03}, \citet{silverman12}, and references therein.  The two-dimensional (2D) spectra were bias-corrected, flat-field corrected, adjusted for varying gains across different chips and amplifiers, and trimmed.  Cosmic-ray rejection was applied using the {\tt pzapspec} algorithm to individual frames.  Multiple frames were then combined with appropriate masking.  One-dimensional (1D) spectra were extracted using the optimal algorithm \citep{Horne86}.  The spectra were wavelength-calibrated using internal comparison-lamp spectra with linear shifts applied by cross-correlating the observed night-sky lines in each spectrum to a master night-sky spectrum.  Flux calibration was performed using standard stars at a similar airmass to that of the science exposures, with ``blue'' (hot subdwarfs; i.e., sdO) and ``red'' (low-metallicity G/F) standard stars.  We correct for atmospheric extinction.  By fitting the continuum of the flux-calibrated standard stars, we determine the telluric absorption in those stars and apply a correction, adopting the relative airmass between the standard star and the science image to determine the relative strength of the absorption.  We allow for slight shifts in the telluric A and B bands, which we determine through cross correlation.  For dual-beam spectrographs, we combine the sides by scaling one spectrum to match the flux of the other in the overlap region and use their error spectra to correctly weight the spectra when combining.  More details of this process are discussed elsewhere \citep{Foley03, silverman12, siebert19}.

Data obtained with ALFOSC and Binospec were reduced using standard techniques, which included correction for bias, overscan, and flat-field. Spectra of comparison lamps and standard stars acquired during the same night and with the same instrumental setting have been used for the wavelength and flux calibrations, respectively. When possible, we further removed the telluric bands using standard stars. Given the various instruments employed, the data-reduction steps described above have been applied using several instrument-specific routines. We employed standard \textsc{IRAF} commands to extract all spectra.

The LRS2 data were processed with \texttt{Panacea}\footnote{\url{https://github.com/grzeimann/Panacea}}, the HET automated reduction pipeline for LRS2.  The initial processing includes bias-correction, wavelength calibration, fiber-trace evaluation, fiber normalisation, and fiber extraction; moreover, there is an initial flux calibration from default response curves, an estimation of the mirror illumination, as well as the exposure throughput from guider images.  After the initial reduction, we used an advanced code designed for crowded IFU fields to perform a careful sky subtraction and host-galaxy subtraction.  Finally, we modelled the target SNe with a \citet{Moffat69} PSF model and performed a weighted spectral extraction.

\begin{figure*}
\centering
\includegraphics[width=\textwidth]{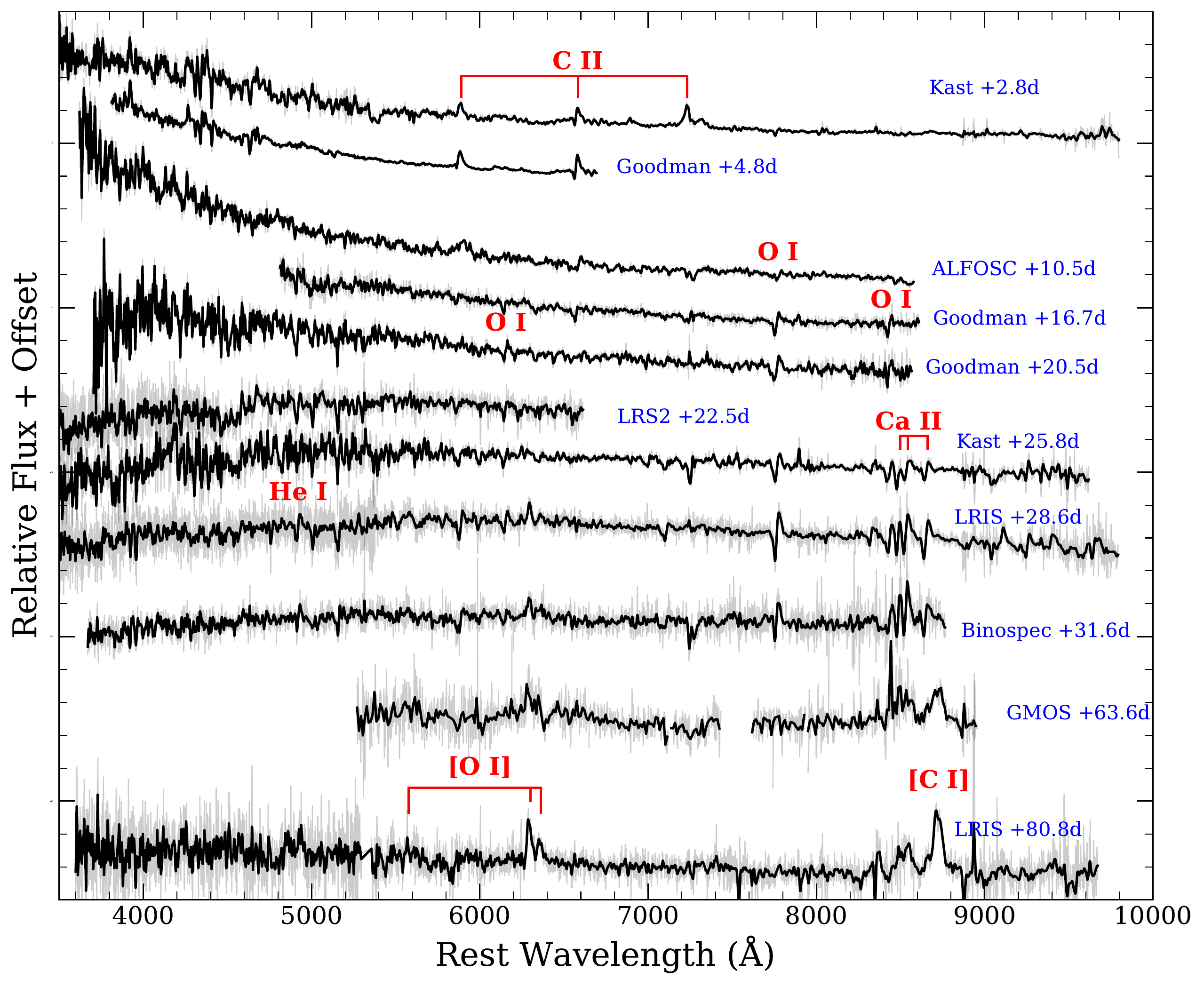}
\caption{Complete set (11) of optical spectra of SN~2022ann taken between +2.7 days and +80.8 days relative to $o$-band maximum brightness (source and phase denoted in blue). Visually prominent lines of C, O, Ca, and He are labeled with red text. Wavelength regions with low signal-to-noise ratio (S/N) have been trimmed. Spectra plotted in black have been smoothed using a Gaussian kernel. Unsmoothed spectra are plotted in gray. \label{fig:spectral_series}}
\end{figure*}

We present our full optical spectral time series of SN~2022ann, consisting of 11 spectra obtained between +2.8 and +80.8~days, in Figure~\ref{fig:spectral_series}.

\begin{figure}
	\includegraphics[width=0.49\textwidth]{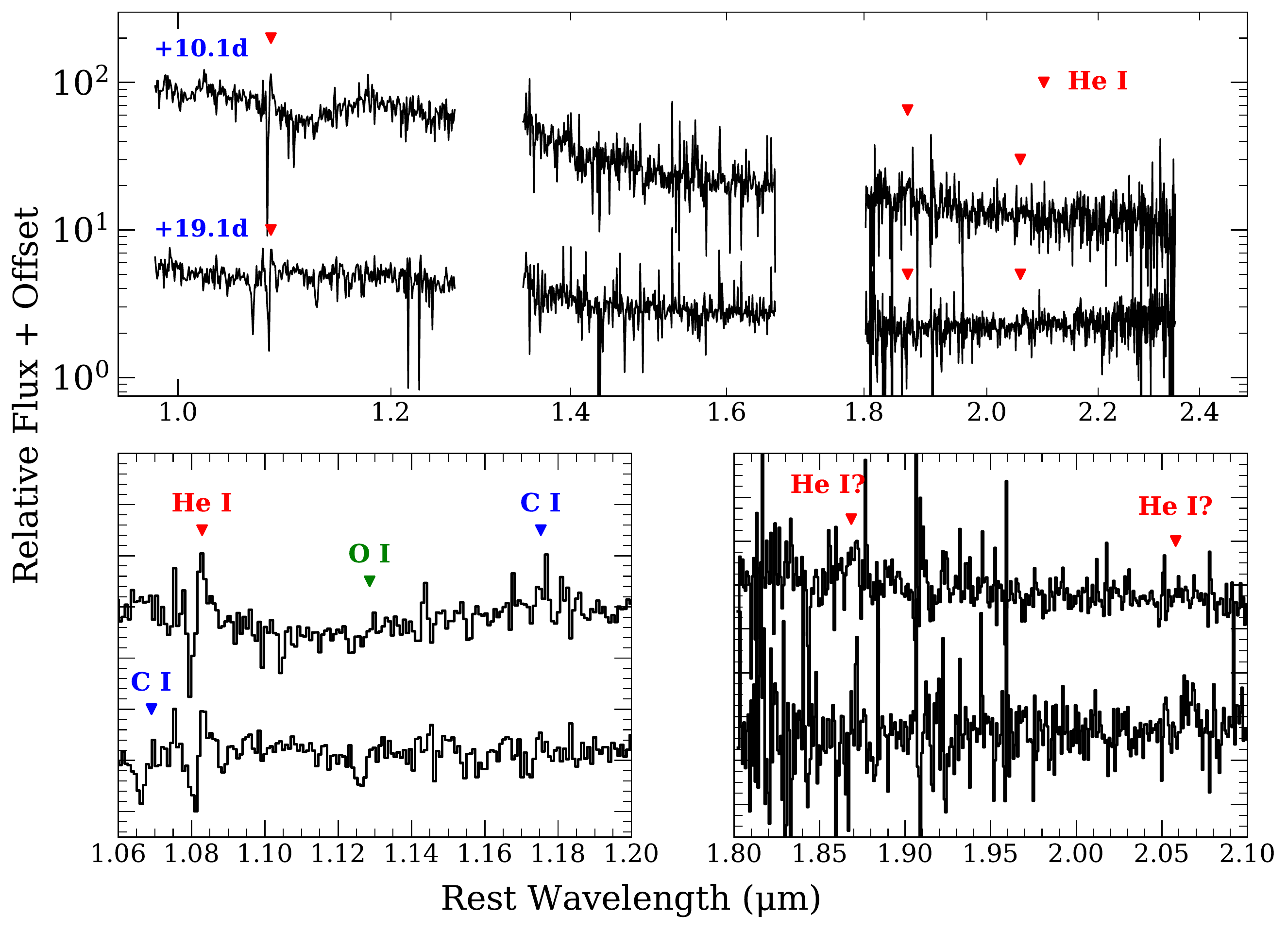}
    \caption{NIR spectra of SN~2022ann taken at +10.1 and +19.1~days relative to $o$-band maximum. Spectra have been binned by a factor of four. Rest wavelengths of several He lines are marked in the top panel by red triangles. The bottom-left panel shows a close-up view of the spectra near prominent features; \ion{He}{I}, \ion{C}{I}, and \ion{O}{I} lines are marked. Note that the \ion{He}{I} 1.083~$\mu$m line is well isolated owing to the small velocity of SN~2022ann.  The bottom-right panel shows a zoom-in on the region of \ion{He}{I} 1.86~$\mu$m and 2.06~$\mu$m; however, we cannot identify any emission because of the relatively low S/N at these wavelengths.}
    \label{fig:nir_series}
\end{figure}

We obtained NIR (0.94--2.45~$\mu$m) spectra of SN~2022ann using the Near-Infrared Echellette Spectrometer \citep[NIRES;][]{NIRES} on the 10~m Keck~II telescope as part of the Keck Infrared Transient Survey (KITS), a NASA Keck Key Mission Strategy Mission Support program (PI R. Foley). A log of observations is given in Table~\ref{tab:speclog}. We observed the SN at two positions along the slit (AB pairs) to perform background subtraction.  An A0V star was observed immediately before or after the science observation. We reduced the NIRES data using \texttt{spextool v.5.0.2} \citep{cushing2004}; the pipeline performs flat-field corrections using observations of a standard lamp and wavelength calibration based on night-sky lines in the science data. We performed telluric correction using \texttt{xtellcor} \citep{vacca2003}.  The NIR spectra are shown in Figure~\ref{fig:nir_series}.

Observations were primarily coordinated using {\tt YSE-PZ} \citep{coulter_d_a_2022_7278430}; an open-source, general-purpose Target and Observation Management (TOM) platform. 

\section{Analysis}\label{sec:results}

Below we examine the spectra and light curves of SN~2022ann, detailing its unique characteristics and comparing to other SESNe. We also construct and model its bolometric light curve, as well as provide an analysis of the host galaxy.

\subsection{Photometric Analysis}\label{sec:photanalysis}

Although the explosion date and rise time are not strongly constrained (see discussion in Section~\ref{sec:observations}), SN~2022ann appears to have a rise time of $\sim$10~days, making it a relatively fast-rising SN, comparable to other SNe~Icn and SNe~Ibn \citep[e.g.,][]{Hosseinzadeh2017}.  The light curves of SN~2022ann are not well sampled in all observed bands right after discovery, so we rely on the $o$-band photometry to characterise the time of maximum brightness and the overall shape of the light curves near peak. We report the luminosity at discovery and maximum brightness in Table \ref{tbl:paramtable}.

SN~2022ann has a similar shape in all $roi$ bands for epochs with overlapping data, and we use the combination of the $o$ and $r$ light curves, which have complementary temporal coverage, to better assess the evolution of SN~2022ann.  In these bands, SN~2022ann is slowly evolving for about 25~days after discovery; it rises by 0.03~mag from discovery to peak (in 2~days) and declines by 0.25~mag in 18~days after peak.  SN~2022ann varies by only 0.5~mag over a period of $>$24 days.  After the relatively consistent brightness (in these bands), the SN begins to decline much faster (0.9~mag~day$^{-1}$) starting around 25~days after maximum.  Both bluer and redder bands seem to have a more pronounced rise and decline around maximum light, but the lack of data in those bands makes it difficult to have strong conclusions about their morphology.

\begin{figure*}
\centering
\includegraphics[width=\textwidth]{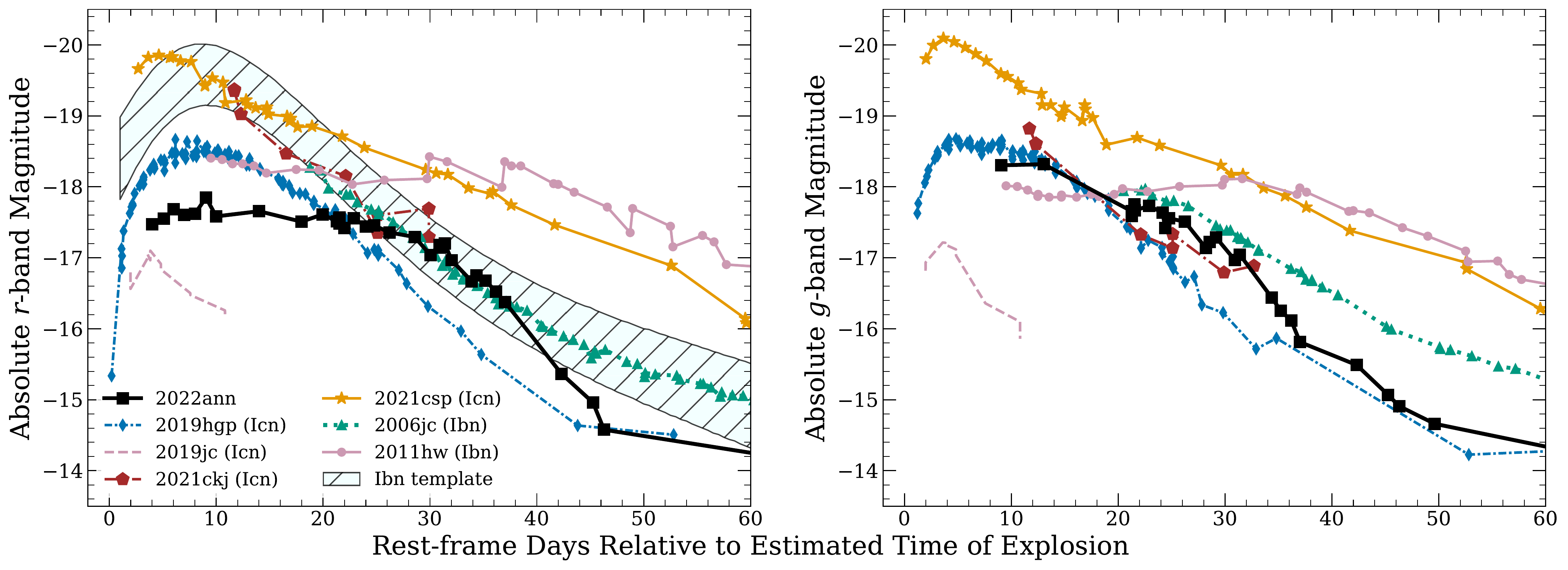}
\caption{Light-curve comparisons in $r$ (left) and $g$ (right) to similar astronomical transients; $o$-band and $c$-band photometry for SN~2022ann is included in the respective comparisons to increase coverage. When $r$ and $g$ photometry for comparison objects was not available in literature, the most similar available filters were selected and noted.  In the left panel, we also display in hatched-blue the \citet{Hosseinzadeh2017} SN~Ibn template light curve. As the explosion time for SN~2006jc is not well constrained, we match the estimated maximum-light epoch presented by \citet{Pastorello07} to the maximum of the SN~Ibn template. SN~2022ann is slow-evolving and somewhat faint relative to the other objects around peak.}
\label{fig:lc_comp}
\end{figure*}

In Figure \ref{fig:lc_comp}, we compare the $o$/$r$-band and $c$/$g$-band light curves of SN~2022ann with those of the other known SNe~Icn, two SNe~Ibn (SNe~2006jc, \citealp{Foley07, Pastorello07}; and 2011hw, \citealp{Smith12}), and the $R$-band SN~Ibn template provided by \citet{Hosseinzadeh2017}. Compared to these SNe, SN~2022ann has a unique evolution in $r$ with its relatively flat and long-lived peak. SN~2022ann declines roughly 0.035~mag~day$^{-1}$ for the first 10 days after peak, while SNe~2019jc and 2019hgp decline at roughly twice that rate. Between 30 and 50 days after peak, SN~2022ann declines faster (0.090~day$^{-1}$) than SN~2019hgp (0.055~day$^{-1}$). 

SN~2022ann has relatively low peak $r$-band luminosity ($M_r {\rm ~peak} = -17.8$~mag), but falls within the bounds of the SN~Icn sample. Despite showing a high degree of uniformity in their early-time spectra (see Section~\ref{sec:specanalysis}), SNe~2022ann, 2019hgp, and 2019jc vary in peak absolute brightness by $\sim$1.5~mag. SN~2022ann has significantly lower luminosity at peak than SNe~2021ckj, 2021csp, and the SN~Ibn template (which \cite{Hosseinzadeh2017} notes is biased toward bright and slowly-evolving objects). SN~2022ann evolves more slowly near its maximum brightness, and declines faster at later times relative to other SNe~Icn and the SN~Ibn template; however, SN~Ibn~2011hw has a similar plateau near peak \citep{Smith12}.

In contrast, the $g$-band evolution of SN~2022ann is similar to that of SN~2019hgp. They reach comparable peak luminosity ($-18.3$ and $-18.7$ mag, respectively), and decline rapidly at 0.09~mag~day$^{-1}$ afterward. Unlike SN~2011hw, which has a plateau in both $r$ and $g$, SN~2022ann declines rapidly in the $g$/$c$ band and has an evolution similar to that of other SNe~Icn at this bluer wavelength.

\begin{figure}
	\includegraphics[width=0.46\textwidth]{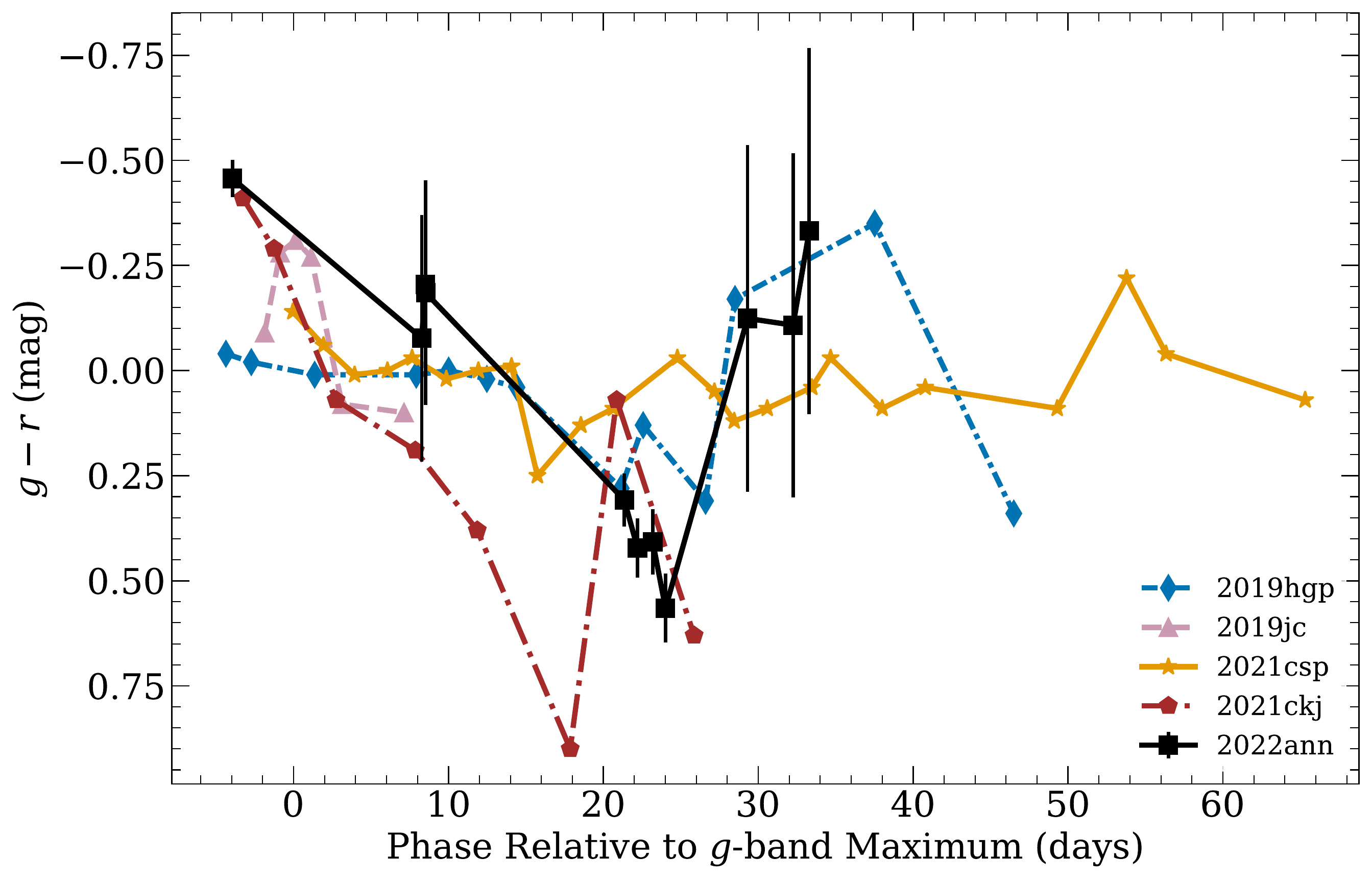}
    \caption{Colour evolution ($g-r$) of SN 2022ann and the SN~Icn sample from \citet{Pellegrino2022}. All SNe~Icn have similar colours.}
    \label{fig:colorcurves}
\end{figure}

We show the $g-r$ colour evolution of SN~2022ann in Figure \ref{fig:colorcurves}, along with the colour curves of the other SNe~Icn presented by \cite{Pellegrino2022}. SN~2022ann begins very blue at maximum light and grows redder over the next 30 days. After this, it quickly becomes blue again. SNe~2019hgp, 2019jc, and 2021ckj also seem to follow this trend; however, the large uncertainties make commenting on the colour-curve morphology difficult.

\subsection{Bolometric Analysis and Modeling}\label{sec:bolanalysis}

In order to understand what is powering SN~2022ann, we construct its bolometric light curve using {\tt extrabol} \citep{ebol_cite}.  {\tt extrabol} interpolates the light curve using a Gaussian process with a 2D kernel, accounting for correlation in both time and wavelength. Each observed epoch is then fit to a blackbody spectral energy distribution (SED), inferring bolometric luminosities, blackbody radii, and blackbody temperatures with time.  Our spectra are well described by a blackbody until +34~days, indicating that this approximation should be appropriate until at least that epoch.   

\begin{figure}
	\includegraphics[width=0.46\textwidth]{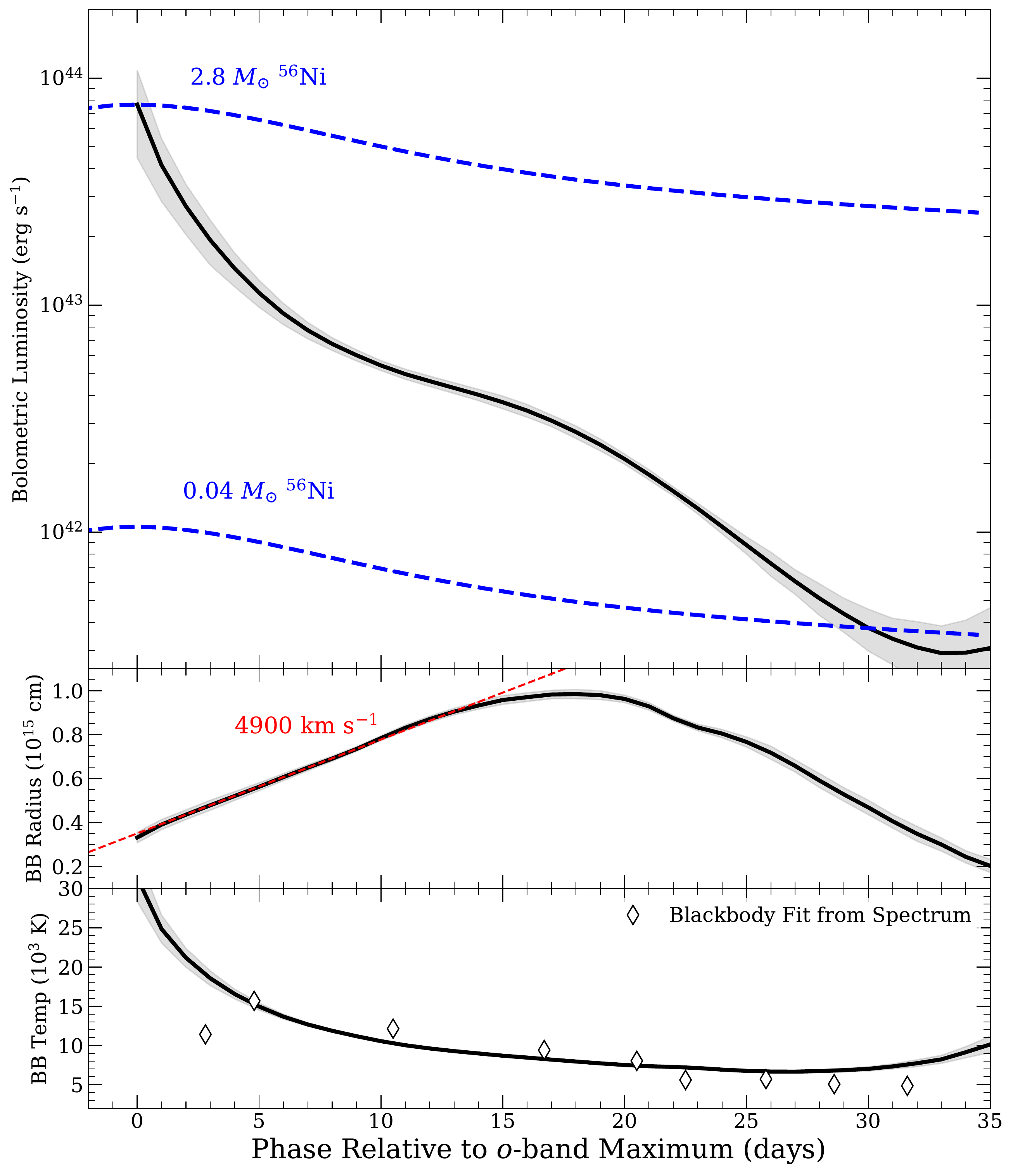}
    \caption{Bolometric light curve (top panel), blackbody radius evolution (middle panel), and blackbody temperature evolution (bottom panel) for SN~2022ann as derived from {\tt Extrabol}.  Measurements are shown as black curves with the 1$\sigma$ uncertainty as a grey band. The temperatures derived from spectra and photometry largely agree, although are discrepant at the earliest spectral epoch. The $g-i$ color at peak is consistent with the lower spectral temperature, but the uncertainty is likely larger than the formal errors shown in the plot. Radioactive-decay models matched to the peak and tail of the light curve (2.8 and 0.04~M$_{\sun}$, respectively) are shown in the top panel in blue. A curve indicating a constant expansion of 4900~km~s$^{-1}$ since the time of explosion, is shown in the middle panel as a dotted red line.}
    \label{fig:lc_bol}
\end{figure}

In Figure~\ref{fig:lc_bol}, we display the resulting bolometric light curve, as well as the derived blackbody radius and temperature evolution for SN~2022ann. We restrict our bolometric light curve to epochs with photometric coverage over at least 4 filters ($\sim$0--35~days after maximum brightness). Soon after explosion, SN~2022ann is very hot ($T \gtrsim 25$,000~K), but cools over the next 15 days until it reaches $\sim$7000~K, where it plateaus. The temperature derived from the photometric SED fitting is consistent with that of fitting the continua of our optical spectra (Section~\ref{sec:specanalysis}), indicating that spectral features are not significantly affecting this analysis. The photosphere initially expands at a rate of 4900~km~s$^{-1}$ until +15~days, when it reaches its maximum extent.  Our +34-days spectrum has a blackbody continuum and P-Cygni features while our +66-day spectrum has prominent forbidden-line emission indicating that the SN has become nebular by that time; therefore, SN~2022ann likely entered its nebular phase between +34 and +66~days, beyond the range of our analysis of the bolometric light curve.

The blackbody properties of SN~2022ann are broadly similar to those of other SNe~Ibn~and~Icn. The temperature evolution of SN~2022ann is similar to that of other SNe~Ibn and Icn \citep{Pellegrino2022}. For the first $\sim$2~weeks after peak luminosity, the blackbody radius of SN~2022ann expands at a constant velocity of 4900~km~s$^{-1}$, which should correspond to the photosphere.  This value is significantly different from the absorption velocity (800~km~s$^{-1}$), further indicating that the spectral features originate from the CSM.  However, we note that we never detect any spectral features with $v > 1000$~km~s$^{-1}$, which is puzzling considering that our last epochs appear to be in the nebular phase.

The photometric properties of SN~2022ann at maximum light are inconsistent with radioactive decay of $^{56}$Ni. If SN~2022ann is powered by $^{56}$Ni, its peak bolometric luminosity of $\sim 10^{44}$~erg~s$^{-1}$ is directly related to the $^{56}$Ni mass \citep{arnett82}.  Using this relation, we determine that SN~2022ann would require 2.8~M$_{\sun}$ of $^{56}$Ni to match its peak luminosity. However, given the rapid decline after maximum and the spectral signatures of CSI present at this epoch (see Section \ref{sec:specanalysis}), it is unlikely that SN~2022ann is powered chiefly by radioactive decay at peak light. In particular, assuming that the SN is optically thin at late times with instantaneous $\gamma$-ray trapping and full efficiency, the late-time luminosity limits the $^{56}$Ni mass to $<$0.04~M$_{\sun}$, smaller than the low end of the SESN population \citep{Lyman16}. We therefore rule out radioactive decay as the primary power source for SN~2022ann near peak.

The narrow spectral features and blue colours at early times suggest that SN~2022ann is primarily powered by CSI. To estimate the SN and CSM properties with the goal of gaining insight into the progenitor system, we use {\tt MOSFiT} \citep{MOSFiT} to fit our photometry with a CSI model \citep[as described by][]{Chevalier1982,villar2017theoretical,chatzopoulos2012generalized,jiang2020extended}. The inner and outer SN ejecta density distributions are modeled as two power laws ($\rho_{\rm inner} \propto r^{-\delta}$ and $\rho_{\rm outer} \propto r^{-n}$, respectively). In our modeling, we assume fixed power laws with $\delta = 1$ and $n = 12$. We assume a fixed inner CSM radius ($R_0$) of 10$^{14}$~cm, a kinetic to thermal energy conversion efficiency ($\epsilon$) of 0.5, an optical opacity ($\kappa$) of 0.34~cm$^{2}$~g$^{-1}$, and minimum temperature of 7000~K. We then allow for the following parameters to vary freely:
\begin{enumerate}
    \item $M_{\rm ej}$, the ejecta mass;
    \item $M_{\rm CSM}$, the CSM mass;
    \item $v_{\rm ej}$, the ejecta velocity;
    \item $\rho$, the density at $R_0$;
    \item $n_{\rm H, host}$, the hydrogen column density of the host galaxy;
    \item $s$, the power-law index in the CSM distribution $\rho_{\rm CSM} \propto r^{-s}$;
    \item $t_{\rm exp}$, time of explosion relative to maximum light; and
    \item $\sigma$, the noise term.
\end{enumerate}

\begin{figure}
	\includegraphics[width=0.48\textwidth]{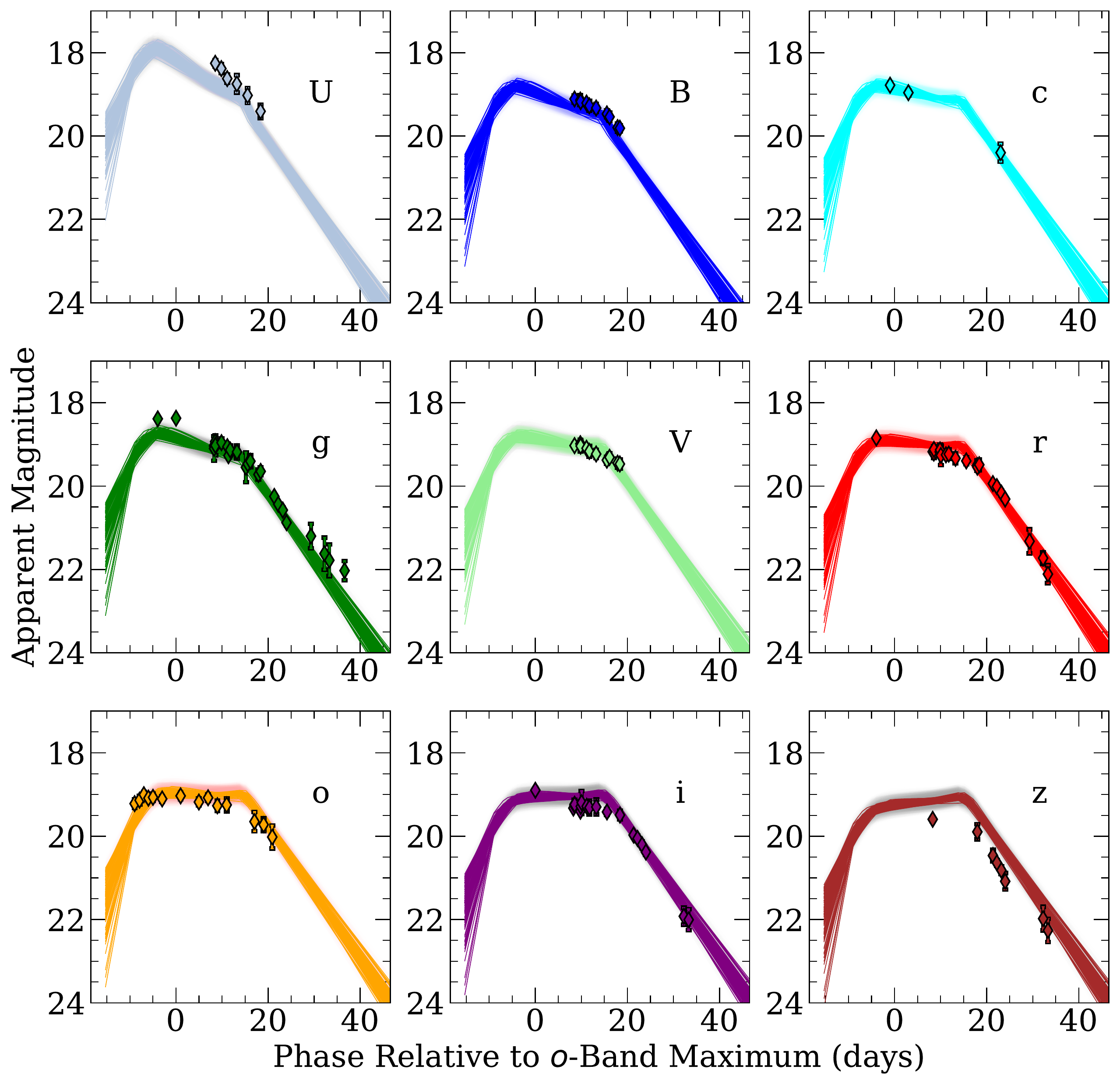}
    \caption{Band-by-band light-curve fits with CSM model from {\tt MOSFiT}. Photometry is plotted with scatter points, while modeled light curves are plotted with solid lines. We show 5000 iterations of the model with {\tt MOSFiT}.}
    \label{fig:csm_fit}
\end{figure}

We show the band-by-band light-curve fit from {\tt MOSFiT} in Figure \ref{fig:csm_fit} and the best-fit CSM model parameters in Table~\ref{tbl:csitable}. SN~2022ann is consistent with 1.73~M$_{\odot}$ of ejecta interacting with 0.19~M$_{\odot}$ of CSM. This ejecta mass is in good agreement with what was modeled for other SNe~Icn by \citet{Pellegrino2022}, and is in disagreement with a massive WR progenitor (pre-SN mass $\gtrsim 10$~M$_{\odot}$). On the other hand, the measured CSM mass and inner CSM density are significantly lower than what was reported for other SNe~Icn by \citet{Pellegrino2022}. These discrepancies likely arise from differing assumptions when modeling the light curves rather than something physically intrinsic to SN~2022ann. We assume $\epsilon = 0.5$ and $R_0 = 10^{14}$~cm, whereas \citet{Pellegrino2022} allowed these parameters to vary freely in their fits and found them to be $\sim$0.03 and $\sim 4\times 10^{14}$ (respectively) in their fits of other SNe~Icn. $\epsilon$ and $R_0$ are degenerate with CSM mass and inner CSM density respectively. Since we assume a much higher energy conversion efficiency, it is not surprising that our fit for SN~2022ann requires far less CSM to reproduce a comparable luminosity. Similarly, the differing inner radius can explain the differences in inner CSM density. Nevertheless, while more rigorous modeling of SNe~Icn is needed to ascertain details about the CSM configuration, our modeling of SN~2022ann shows that it is inconsistent with a WR progenitor.

Our fit also gives an ejecta velocity of $\sim$4500~km~s$^{-1}$, consistent with the photospheric velocity measured from the bolometric light curve (4900~km~s$^{-1}$). This velocity is significantly lower than the ejecta velocities measured for the other SNe~Icn; 4500~km~s$^{-1}$ is nearly half that of the next-lowest velocity (SN~2019jc) and $\sim$25\% of the highest-velocity SNe~Icn (SNe~2019hgp and 2021csp). This again raises the question of why spectral features with $v > 1000$~km~s$^{-1}$ are never seen.

\begin{table}
    \centering
    \caption{CSM model parameters for SN~2022ann from 5000 iterations in {\tt MOSFiT}.} 
     \begin{tabular}{lr}
\hline
\hline
\vspace{0.2cm}
$M_{\rm ej}$ (M$_{\odot}$) &  $1.73^{+0.20}_{0.16}$\\ 
\vspace{0.2cm}
$M_{\rm CSM}$ (M$_{\odot}$) &  $0.19^{+0.018}_{-0.018}$\\ 
\vspace{0.2cm}
$v_{\rm ej}$ (km~s$^{-1}$) &  $4460^{+200}_{-200}$\\
\vspace{0.2cm}
$\log \rho$ (g~cm$^{-3}$)  & $-12.03^{+0.17}_{-0.20}$ \\
\vspace{0.2cm}
$\log n_{{\rm H, host}}$ (cm$^{-2}$) & $17.54^{+1.01}_{1.04}$\\
\vspace{0.1cm}
$s$ & $0.35^{+0.26}_{-0.26}$\\
\vspace{0.2cm}
$t_{\rm exp} ({\rm days})$ & $-10.01^{+1.39}_{-1.23}$\\
\vspace{0.2cm}
$\log \sigma$ & $-0.65^{+0.05}_{-0.02}$\\
\hline
\end{tabular}\label{tbl:csitable}
\end{table}


\subsection{Spectroscopic Analysis}\label{sec:specanalysis}

\begin{figure*}
\centering
\includegraphics[width=\textwidth]{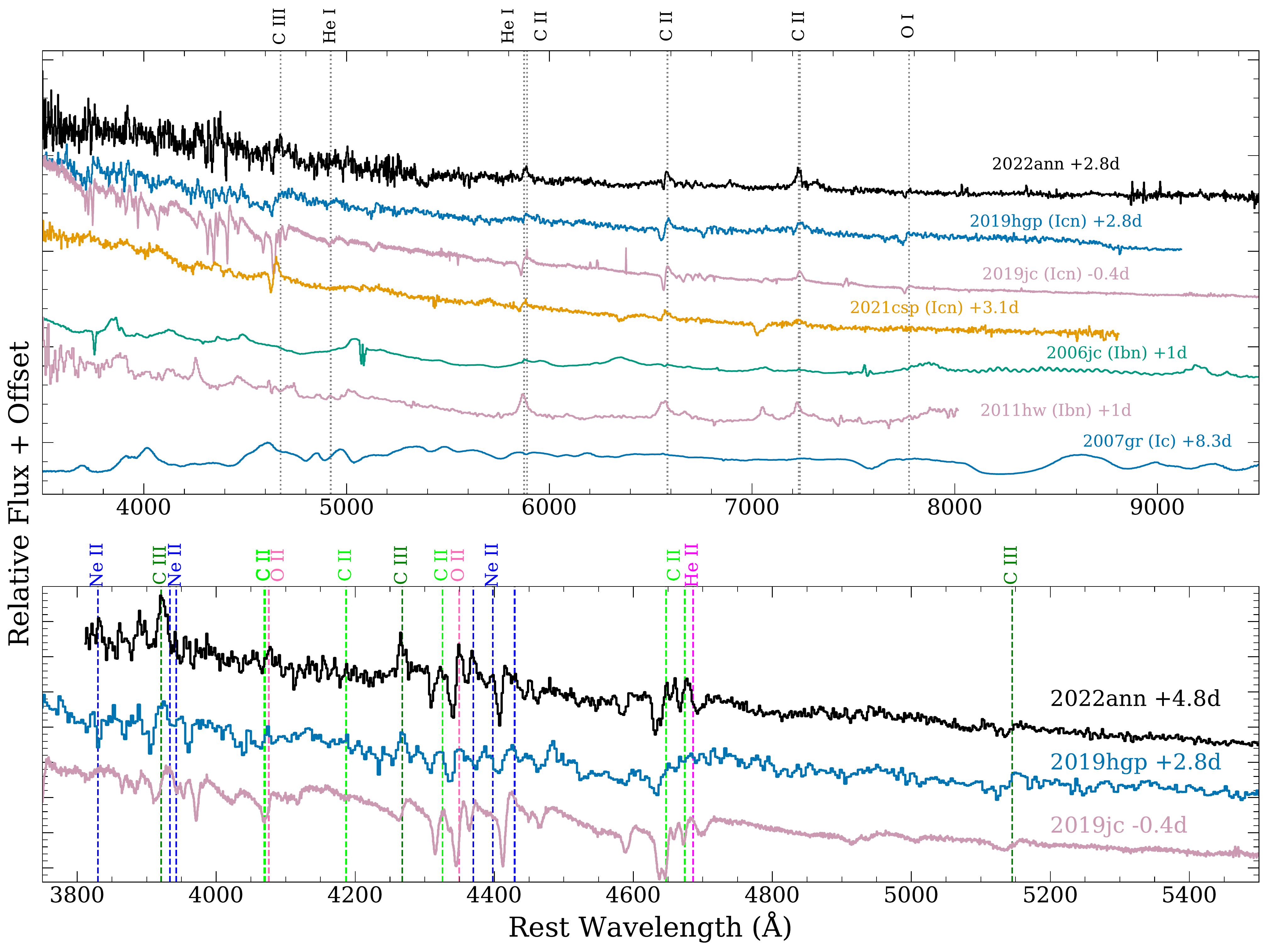}
\caption{Early-time ($<$10~days after peak light) spectral comparison of SN~2022ann and other SNe~Icn, SN~Ibn~2006jc, SN~Ibn~2011hw, and SN~Ic~2007gr. Labeled phases are with respect to $r$- or $o$-band maximum light. Notable features are marked with vertical lines. The bottom panel displays a subset of the spectra (SNe~2019hgp, 20219jc, and 2022ann) in the region $<$5200~\AA. We mark several potential features from highly-ionized lines of C, O, and Ne. We mark \ion{He}{II} in magenta, but it does not match the nearby features in SN~2022ann particularly well. Spectra for SNe~2006jc, 2007gr and 2019hgp were obtained from {\tt WISeREP} \citep{wiserep}. \label{fig:spec_compare_early}}
\end{figure*}

The earliest spectrum in our dataset is also the spectrum with which we classified SN~2022ann \citep{Davis2022}.  As noted in that report, the early-time spectrum has a blue continuum with prominent narrow \ion{C}{II} and \ion{C}{III} P-Cygni features.  There is also a weak \ion{O}{I} $\lambda$7774 P-Cygni line. The minimum of the absorption for the \ion{C}{II} $\lambda$6583 line has a velocity of $-870$~km~s$^{-1}$.  This spectrum is similar to that of other SNe~Icn soon after explosion \citep{Pellegrino2022}, with SN~2022ann being especially similar to SN~2019hgp \citep{Gal-Yam2022} at a comparable phase (Figure~\ref{fig:spec_compare_early}).

Our spectral series, shown in Figure~\ref{fig:spectral_series}, exhibits a unique evolution, even among SNe~Icn.  The continuum for each spectrum is smooth and can be described by a blackbody.  Until +31.6~days, the spectra have narrow P-Cygni features from \ion{C}{II} ,\ion{O}{I}, and later \ion{Ca}{II} and \ion{He}{I}. The earliest spectra ($\lesssim$4.8~days) also show a forest of highly ionized lines of C, O, and Ne blueward of $\sim$5200\AA, which are shown in Figure \ref{fig:spec_compare_early}. The later spectra continue to have narrow carbon and oxygen emission, although as forbidden [\ion{C}{I}] and [\ion{O}{I}].  This transition from permitted P-Cygni profiles to pure-emission forbidden lines between +31.6 and +63.6~days marks the time when the CSM becomes optically thin. This period also corresponds to when the light curve begins to fade quickly (see Figure~\ref{fig:photometry}).

The spectral sequence displays a gradual weakening of high-ionization lines with a contemporaneous strengthening of low-ionization lines (both relative to the continuum). Between +2.8 days and +16.7 days, we identify several narrow P-Cygni lines from \ion{C}{II} $\lambda\lambda$5890, 6583, 7231, 7236, \ion{C}{III} $\lambda$4650, and \ion{O}{I} $\lambda$7774. We identify several transitions from \ion{C}{II}, \ion{C}{III}, \ion{O}{II}, \ion{Ne}{I}, and \ion{Ne}{II} in the forest of lines between 3400~and~5200~\AA\ in our earliest spectra. At these epochs, \ion{He}{I} features are not detected in this wavelength range. There is faint emission near the expected location of \ion{He}{II} $\lambda$4686, but the peak is blueshifted by $\sim$600~km~s$^{-1}$; the offset may be the result of blending with other lines.  Regardless, we do not unambiguously detect any He features in these early-time optical spectra.

In the +10.1-day NIR spectrum, we detect the 1.083~$\mu$m \ion{He}{I} line with a P-Cygni profile and an absorption velocity of $-720$~km~s$^{-1}$.  We do not detect any He lines in a contemporaneous (+10.5-day) optical spectrum.  The NIR He feature becomes stronger in the +19.1-day spectrum, while we still do not detect any He lines in a contemporaneous (+20.5-day) optical spectrum.  We finally detect \ion{He}{I} $\lambda$4922 in the +28.6-day optical spectrum.

\begin{figure*}
\centering
\includegraphics[width=\textwidth]{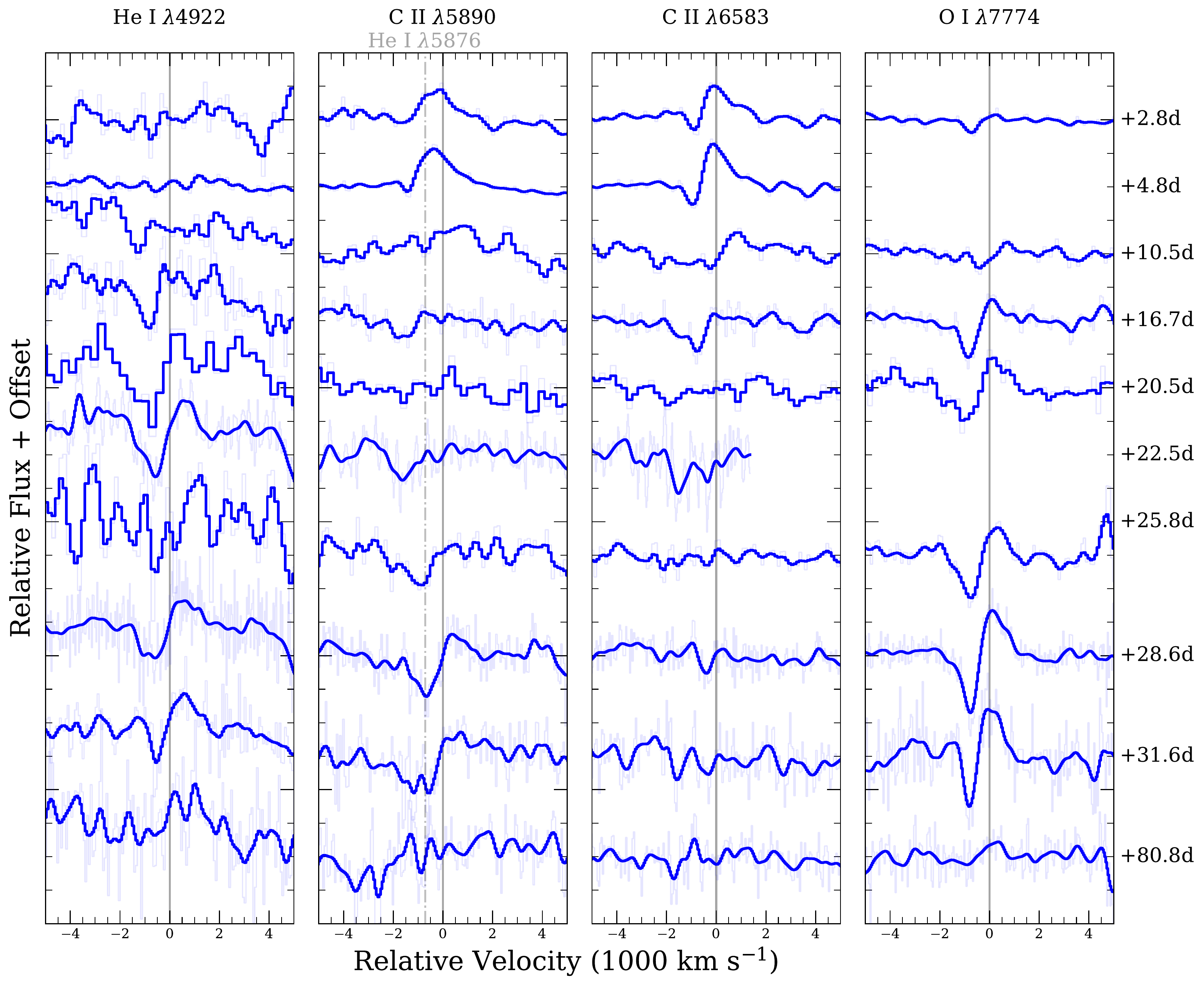}
\caption{Evolution of prominent lines of He, C, and, O in SN~2022ann plotted on a velocity scale relative to their rest wavelengths. From left to right, the panels show \ion{He}{I} $\lambda$4922, \ion{C}{II} $\lambda$5890, \ion{C}{II} $\lambda$6589, and \ion{O}{I} $\lambda$7774. Solid vertical lines mark rest-frame wavelengths of the previously named lines, while dashed vertical lines mark rest wavelengths of nearby lines that may contribute to the overall line profile. Gaussian-smoothed spectra are plotted in solid blue on top of unsmoothed spectra in faint blue. Phases relative to $o$-band maximum light are denoted on the right.}
\label{fig:feature_series}
\end{figure*}

We zoom in on prominent lines of C, O, and He in Figure \ref{fig:feature_series}. The absorption-component velocities of each line consistently measure at $\sim$800~km~s$^{-1}$, with the notable exception of \ion{C}{II} $\lambda$5890, which is higher at $\sim$1300~km~s$^{-1}$. We suspect that this is caused by contamination from \ion{He}{I} $\lambda$5876 emission, changing the line profile. As the spectrum evolves, the \ion{C}{II} emission is still prominent, and the absorption velocity decreases to $\sim$800~km~s$^{-1}$.

\citet{Pellegrino2022} examined optical spectra of four SNe~Icn and did not detect \ion{He}{I} in any of them.  They detected weak \ion{He}{II} $\lambda$4686 in a single spectrum of one SN (SN~2019jc); however, the spectrum has many absorption lines near that wavelength and the claimed feature is weaker than several other unidentified absorption features.  Their dataset only contained optical spectra and the latest spectrum was obtained 22~days after peak light.  If SN~2022ann had a similar dataset, it also would not have shown a clear He detection.  \citet{Perley2022} detected a weak \ion{He}{I} $\lambda$5876 line in their spectra of SN~2021csp (this SN was included in the \citealt{Pellegrino2022} sample), but the detection required high-S/N data. Furthermore, it is possible that any \ion{He}{I} $\lambda$5876 line in other SNe~Icn is obscured by the absorption component of the adjacent, strong \ion{C}{II} $\lambda$5890 P-Cygni line. It is possible that all SNe~Icn have helium, but either high-S/N optical spectra, NIR spectra, or optical spectra taken $\sim$25 days after peak light are necessary to robustly detect the lines.

As noted from its photometry, SN~2022ann is very blue near peak brightness and quickly becomes redder (see Section~\ref{sec:photanalysis}).  Our first spectrum ($t = +2.8$~days) exhibits a blue continuum with relatively weak spectral features; fitting to the continuum of the spectrum, we find that it is well described by a blackbody with $T  \approx 12,000$~K.  Over time, the continuum becomes redder, but is still well described by a blackbody through our +31.6-day spectrum when the temperature has decreased to $\sim$9000~K.  This evolution is consistent with the temperatures derived from the photometry.

\begin{figure}
	\includegraphics[width=0.48\textwidth]{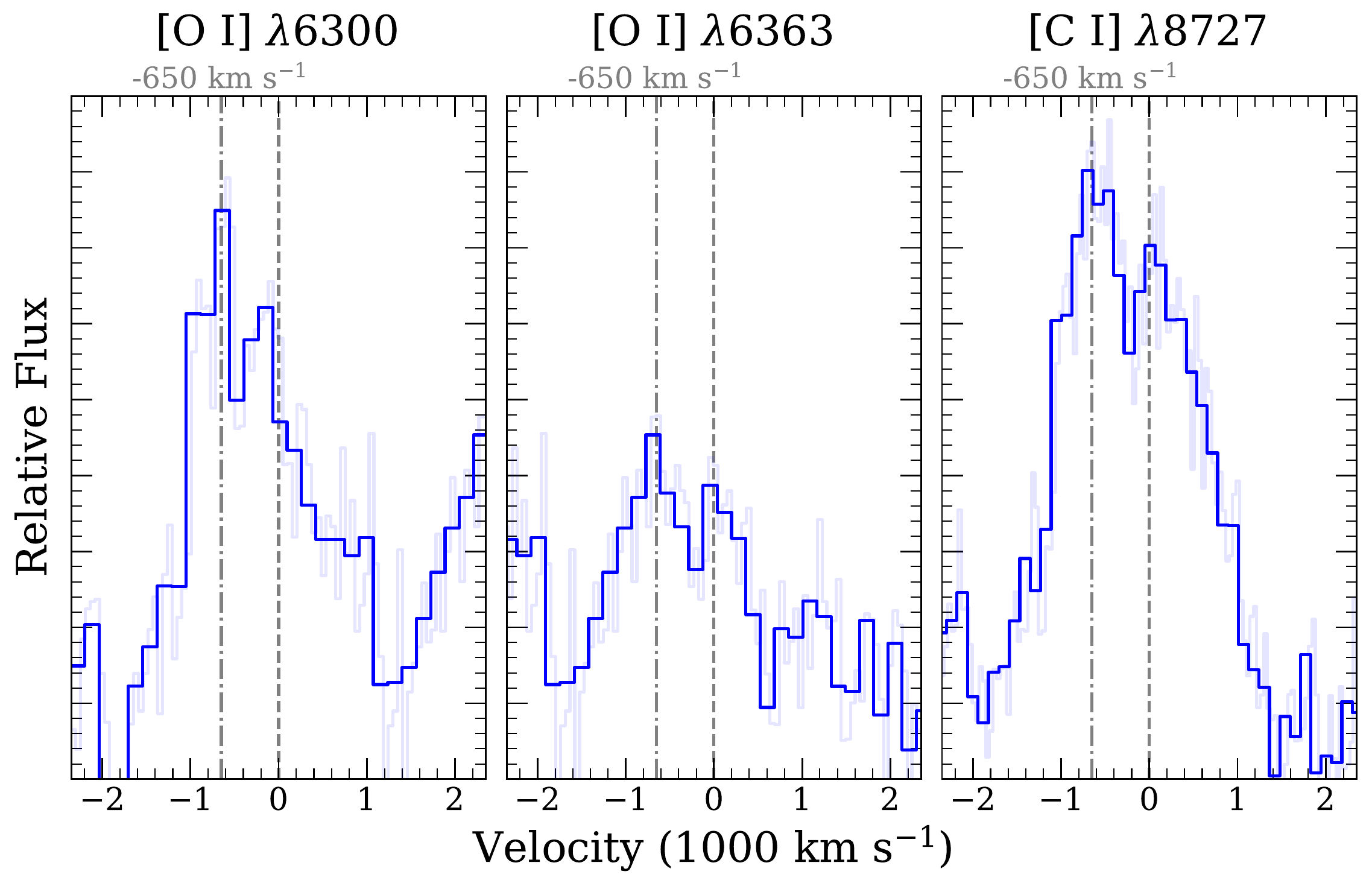}
    \caption{Double-peaked nebular lines of forbidden C and O seen in the LRIS spectrum at +80.8~days plotted in velocity space relative to their rest-frame wavelengths. Unbinned spectra are shown in faded blue lines, while solid blue lines show spectra binned by a factor of 3. Vertical lines show $-650$ and 0~km~s$^{-1}$.}
    \label{fig:nebular_lines}
\end{figure}

Starting around 30~days after peak, SN~2022ann starts to fade quickly.  The +63.6-day and +80.8-day spectra do not have any detected P-Cygni absorption features and we detect forbidden emission lines. Therefore, we consider SN~2022ann to have entered the nebular phase between +31.6 and +63.6~days.  At +80.8~days, the strongest features are from [\ion{O}{I}] and [\ion{C}{I}] (shown in detail in Figure~\ref{fig:feature_series}).  We measure a full width at half-maximum intensity (FWHM) of $\sim$1000~km~s$^{-1}$ for [\ion{O}{I}] $\lambda$6300, consistent with the P-Cygni absorption velocity from earlier epochs and significantly smaller than the photospheric velocity derived from modeling the light curve. 
Thus, these forbidden lines likely originate from the CSM and not the SN ejecta. Shown in Figure \ref{fig:nebular_lines}, the nebular lines are also double-peaked with a blueshifted component at $\sim$-650~km~s$^{-1}$, indicating an asymmetric distribution of CSM. We do not detect broad ($\gtrsim$1000~km~s$^{-1}$) spectral features at any epoch, and so we do not clearly detect any spectral features from the SN ejecta.

In Figure \ref{fig:spec_compare_early}, we compare the +2.8 and +4.8 day spectra of SN~2022ann to those of other SNe at similar epochs ($\sim +1$ week). We include other SNe~Icn, two SNe~Ibn \citep[SNe~2006jc and 2011hw;][]{Foley07, Pastorello07, Smith12}, and the normal SN~Ic~2007gr \citep{Valenti08}. At early times ($\sim +2.8$ days relative to maximum light), SN~2022ann is similar to other SNe~Icn, which together show a high degree of homogeneity despite the scatter in their luminosities. Particularly, we find excellent matches to SNe~2019hgp~and~2019jc, all of which have blue continua, prominent \ion{C}{II} and \ion{O}{I} P-Cygni features, and a similar forest of lines blueward of $\sim$5200~\AA. Notably, SNe~2019hgp, 2019jc, and 2022ann all contain \ion{Ne}{II}. This species is not commonly seen in CCSNe, yet three of the five SNe~Icn \citep[SNe~2019jc, 2019hgp, and now 2022ann;][]{Gal-Yam2022, Pellegrino2022} have \ion{Ne}{II} in their early-time spectra. While SN~2021csp has \ion{C}{II} lines similar to these SNe~Icn, \ion{O}{I} is much weaker. The forest of highly-ionized lines C, O, and Ne observed in the other SNe~Icn is not present in its spectra, though this may be due to a high density of Fe lines saturating the continuum in that region.

The SN~Ibn spectra are qualitatively similar to those of SNe~Icn, with analogous narrow lines of He and blue continua. The SNe~Ibn in our comparison have a combination of narrow emission lines with and without P-Cygni absorption components which arise from different regions in the system, or from viewing-angle dependencies on the CSM. Meanwhile, the SNe~Icn have exclusively P-Cygni line profiles. At this epoch, none of the SNe~Icn show broad features as seen in SN~2007gr.

\begin{figure*}
\centering
\includegraphics[width=\textwidth]{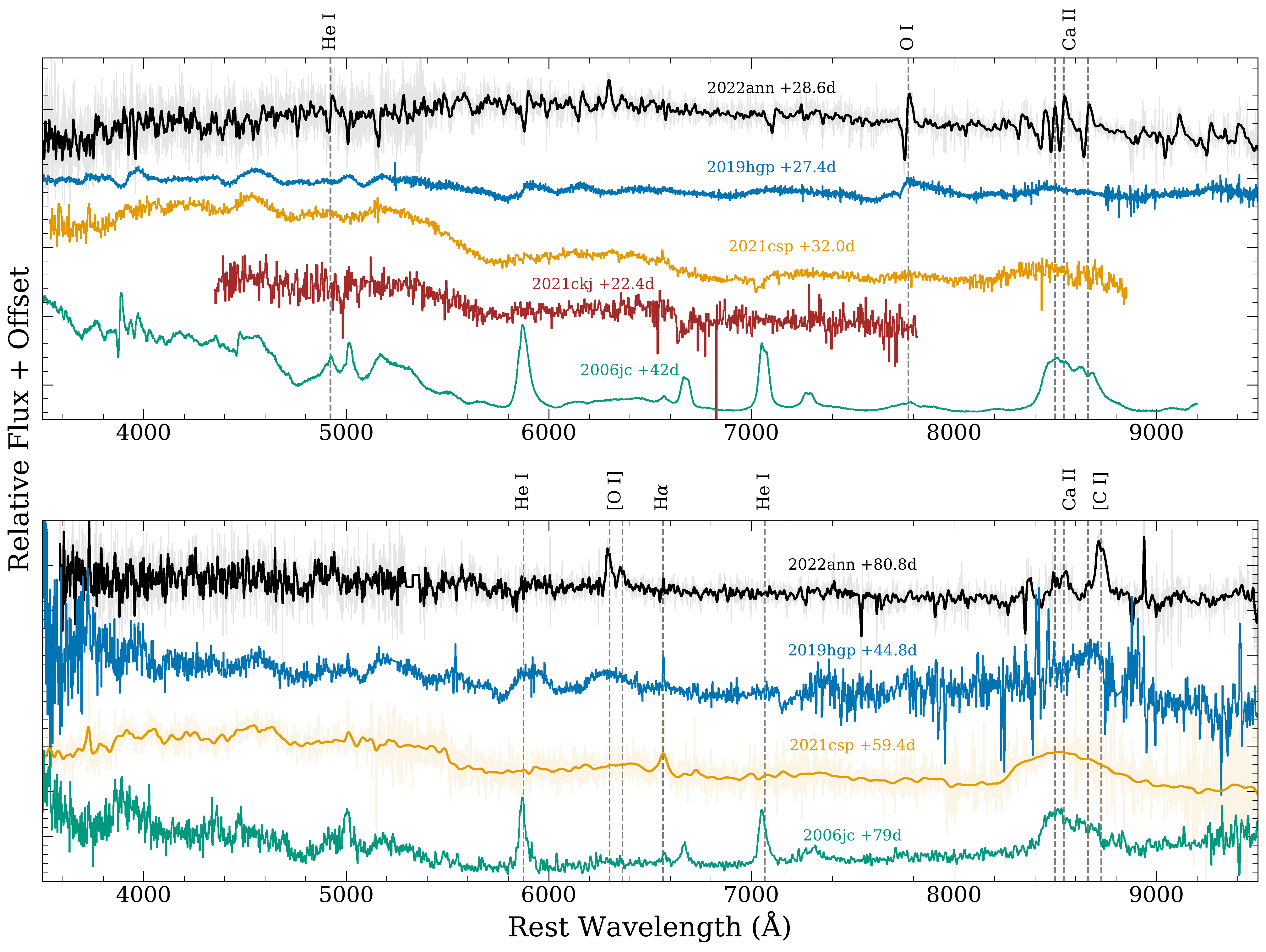}
\caption{Spectral comparison of SN~2022ann other SNe~Icn~and~Ibn around a month after maximum light (top panel) and during the nebular phase (bottom panel). Notable features are marked with vertical dotted lines. Phases are shown relative to maximum. The latest available spectra were chosen for SNe~2019hgp and 2021csp. A month after maximum, SN~2022ann is the only SN~Icn that retains its narrow emission lines and does not show broad ejecta features. In the nebular phase, it still exhibits only intermediate-width lines of forbidden C and O, quantitatively similar to the nebular He lines in SN~2006jc.
\label{fig:spec_compare_late}}
\end{figure*}

We compare these SNe after peak and at late times (if data are available) in Figure \ref{fig:spec_compare_late}. After their similar spectral behaviour at peak light, the SNe~Icn diverge from one another around one month after peak.  As discussed by \citet{Pellegrino2022} at these times, the luminous SNe~2021csp and 2021ckj have developed broad features with high velocities ($\sim$10,000~km~s$^{-1}$) indicative of SN ejecta. The spectra also show a ``break'' in their continua blueward of 6000~\AA, which \citet{Perley2022} attributed to \ion{Fe}{II} fluorescence caused by CSI in the post-shock CSM, a phenomenon originally observed in SNe~Ibn \citep{Foley07}. SN~2019hgp retains narrow lines for longer, but also develops broad P-Cygni features and resembles a spectroscopically-normal SN~Ic \citep{Gal-Yam2022}. SN~2019hgp also develops a similar, albeit weaker, blue continuum break. Uniquely, the spectrum of SN~2022ann does not show any broad features one month after peak, and remains dominated by narrow P-Cygni lines originating from the CSM. SN~2022ann does not develop the blue pseudocontinuum seen in the other SNe~Ibn/Icn.

In the nebular phase, SN~2022ann has a unique spectrum with a relatively flat (in $f_{\lambda}$) continuum and narrow [\ion{C}{I}] and [\ion{O}{I}] emission lines.  In contrast, the other SNe~Icn have blue spectra with an Fe-fluorescence bump, broad undulations, and weak or no narrow emission lines. The emission lines in SN~2022ann are similar in strength and width to the He emission seen in SN~2006jc, indicating that they are from CSM and not SN ejecta. On the other hand, SN~2006jc has a Fe-fluorescence bump and continuum shape similar to the SNe~Icn besides SN~2022ann. SN~2006jc also has an additional thermal component rising in the red end of the optical, attributed to hot dust \citep{Smith07}.

\subsection{Host-galaxy Analysis}\label{sec:host}

SN~2022ann was discovered in a faint host galaxy (SDSS J101729.72--022535) with no published redshift or distance information \citep{Davis2022}. The redshift measured from the CSM lines in the SN spectrum was $z=0.049$. However, the absorption components of these P-Cygni lines can inconsistently shift the wavelength of the maxima. Therefore, we look for emission lines from the host galaxy in the last SN spectrum (when the SN emission is faintest) to obtain a better redshift estimate.  We identify an H$\alpha$ emission line ($\sim 1.7\sigma$) at $z = 0.04938 \pm 0.0004$, which makes SN~2022ann the second-closest known SN~Icn after SN~2019jc \citep{Pellegrino2022}. At this redshift, the galaxy has an absolute magnitude of $M_{r} = -14.4 \pm 0.2$~mag\footnote{Throughout the host analysis, as in the rest of the paper, we assumed $\Lambda$CDM cosmology (H$_{0}$ = 70~km~s$^{-1}$~Mpc$^{-1}$, $\Omega_{M} = 0.3$, $\Omega_{\Lambda} = 0.7$)}, making it a dwarf galaxy with a luminosity intermediate between the Sagittarius Dwarf and the Small Magellanic Cloud, and among the lowest-luminosity SN host galaxies yet discovered \citep{Gutierrez2018, Schulze2021, Taggart2021}. We also note that SN~2022ann appears to be in a galaxy group, since several other galaxies such as V1CG 662 \citep{Lee2017} are at a redshift similar to that of SN~2022ann (Fig.\ref{fig:finder}). 

To characterise the properties of the host galaxy in more detail, we performed elliptical aperture photometry on the host using images from wide-field public surveys. The host was only detected in optical public imaging (it was not detected by \textit{GALEX}, 2MASS, or \textit{WISE}). Host-galaxy photometry (AB mag, not corrected for Galactic extinction) is presented in Table~\ref{tab:hostphotometry}.

We modelled the broad-band SED using the \textsc{Le PHARE} package \citep{Arnouts1999MNRAS, Ilbert2006}, correcting the photometry for Milky Way foreground extinction prior to fitting. We omitted the $u$- and $y$-band limits since they do not usefully constrain the SED models. The code utilises the population-synthesis templates of \citet{Bruzual2003MNRAS}, summed according to an exponentially declining burst of star formation and with stellar metallicities between $0.2\,Z_{\odot} < Z < Z_{\odot}$ and assuming a Chabrier initial-mass function \citep[IMF;][]{Chabrier2003}.  Dust attenuation in the galaxy is applied to the SED models by adopting the \citet{Calzetti2000ApJ} attenuation law. 

\begin{figure}
	\includegraphics[width=0.48\textwidth]{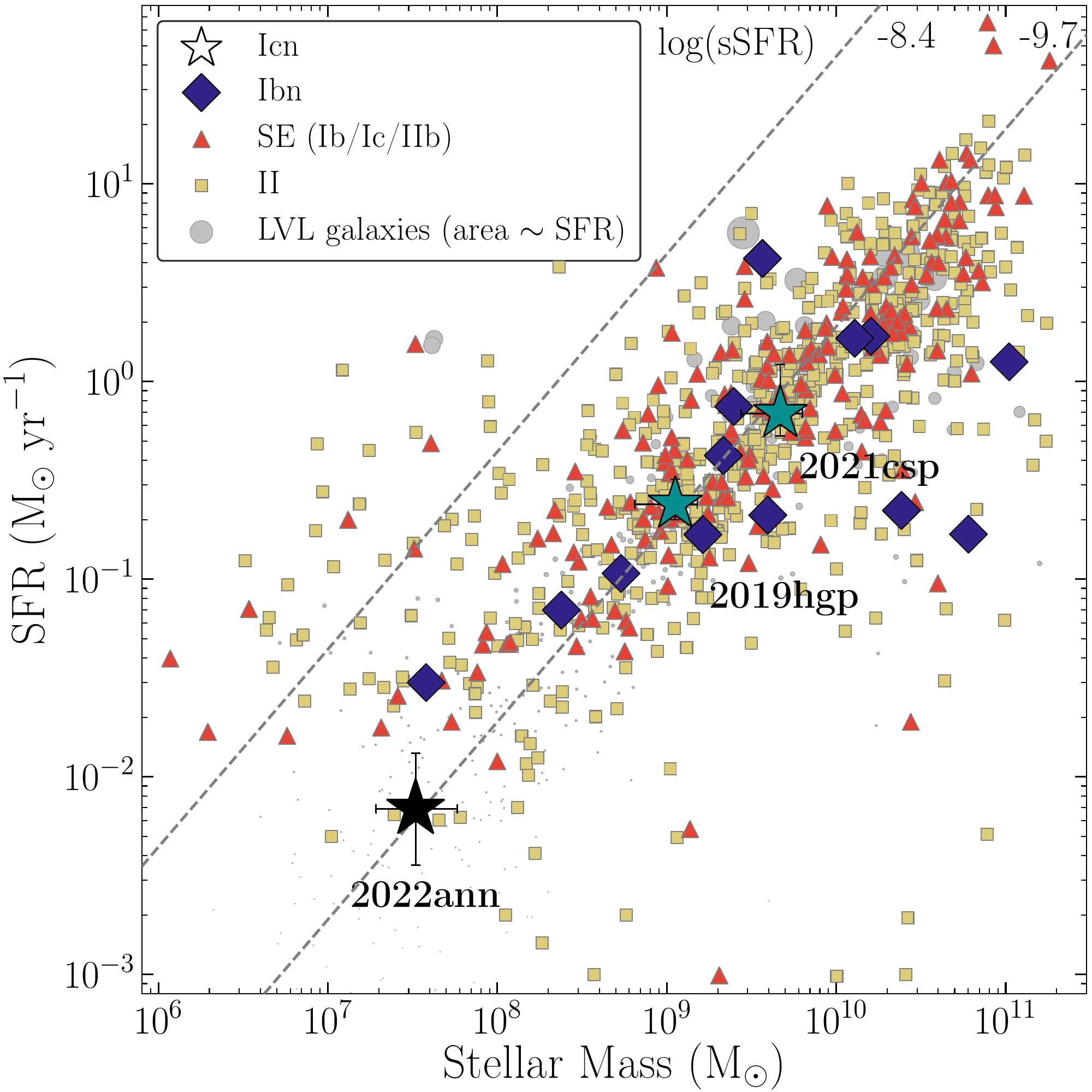}
    \caption{SFR versus stellar mass. Local Volume ($<11$~Mpc) Legacy Survey galaxies are plotted as grey circles \citep[area is scaled in proportion to each galaxy's SFR;][]{Dale2009}. SNe~Ibn (dark blue), SESNe (red triangles), and SNe~II (yellow squares) from ASAS-SN and PTF \citep{Taggart2021,Schulze2021} are plotted. The host galaxies of SNe~Icn  (green stars; Taggart et al., in prep.) that have published physical parameters are labeled with their corresponding SN \citep{Gal-Yam2022,Perley2022}, including SN~2022ann (black star). The SN~Icn host galaxies with published physical parameters are labeled with their corresponding SN \citep{Gal-Yam2022,Perley2022}. Lines of constant sSFR are plotted. The line with log(sSFR) $= -9.7$ is the median sSFR of all comparison SNe, whereas the line at log(sSFR) $= -8.4$ is the median sSFR for galaxies with a stellar mass $< 10^8$~M$_\odot$. The majority of CCSN host galaxies (including SN~Icn hosts) lie on the star-forming galaxy main sequence. However, the host galaxy of SN~2022ann is lower in stellar mass and SFR than the vast majority of galaxies that produce CCSNe.}
    \label{fig:samplebreakdown}
\end{figure}

We derived a stellar mass of log$(M/{\rm M}_\odot) = 7.34^{+0.11}_{-0.30}$ and an integrated SFR of $\log({\rm SFR}) = -2.20^{+0.30}_{-0.09}$~M$_{\odot}$~yr$^{-1}$, both of which are consistent with known local ($<11$~Mpc) dwarf galaxies \citep{Dale2009} --- see the local Volume Legacy galaxies (LVL) plotted in Figure~\ref{fig:samplebreakdown}.  Using the mass and SFR estimates from the SED modelling, we estimate the host-galaxy metallicity, employing the mass-metallicity relation of  \citet{Mannucci2010}, which is roughly continuous even in this low stellar mass regime \citep{Kirby2013}. We derive a metallicity of log$(Z/{\rm Z}_\odot) = 0.10^{+0.05}_{-0.07}$, assuming a solar relative oxygen abundance of 12 + log(O/H) = 8.69 \citep{Asplund2009}.

Figure~\ref{fig:samplebreakdown} shows the integrated SFR plotted against stellar mass for various types of CCSNe from PTF \citet{Schulze2021} and ASAS-SN \citet{Taggart2021}. We see that SN~2022ann has the lowest mass and SFR of any SN~Icn yet discovered (although there are only 5 SN~Icn host galaxies) and is lower in stellar mass and SFR than the vast majority of galaxies that produce CCSNe (both at 3rd percentile for CCSNe).

However, similarly to other SN~Icn host galaxies, SN~2022ann closely follows the star-forming main sequence, with a specific SFR of log(sSFR) = $-9.54^{+0.60}_{-0.20}$, 0.2~dex above the median for CCSN hosts. However, when considering only dwarf galaxies ($M<10^{8}$~M$_{\sun}$), SN~2022ann has a comparatively low sSFR (5th percentile).\footnote{We caution that since we lack UV data, the uncertainties in the derived SFR are large \citep{Childress2013}. A more detailed spectroscopic comparison of SN~2022ann with other SN~Icn host galaxies will be discussed in future work (Taggart et~al., in prep.).)}

\section{Discussion}\label{sec:disc}

SN~2022ann is part of the small but quickly growing class of SNe~Icn.  Its unique photometric and spectroscopic behaviour among the known members demonstrates that this class is relatively heterogeneous.  The detection of He in the spectra of SN~2022ann, and the way in which it was detected, suggests that perhaps all SNe~Icn have He in their systems.  Finally, the extreme host-galaxy properties of SN~2022ann combined with its rare observables constrain its progenitor system.  We discuss these areas in detail below.

\subsection{Helium in SNe~Icn}\label{sec:He}
Optical spectra of SNe~Icn, by definition, lack He emission. If He lines are present at all, they are weak. This is in contrast to SNe~Ibn where He emission lines dominate their spectra. While this difference may be the result of different abundances, it is possible that SNe~Ibn are able to excite He in their circumstellar environments from nonthermal photons while SNe~Icn are not. Detailed modeling of SNe~Ibn and SNe~Icn is beyond the scope of this paper, but considering the detection of He in SN~2022ann, we remark on its ubiquity in SNe~Icn below.

Because of its high ionization and excitation energies, it is generally difficult to detect He in SNe~Ic \citep{Hachinger2012, Desart20, Teffs2020, Williamson2021, Shahbandeh2022}, especially at optical wavelengths.  A number of obstacles could complicate identification of He in SNe~Icn and SN~2022ann in particular. The strongest \ion{He}{I} line in the optical, \ion{He}{I} $\lambda$5876, is coincident with the absorption minimum of the P-Cygni profile of the strong \ion{C}{II} $\lambda$5889 line, making it challenging to disentangle (see Section \ref{sec:specanalysis}). The density of high-ionization C, O, and Ne lines at $\lambda < 5200$~\AA\ in the early-time spectra similarly makes unambiguously identifying \ion{He}{II} $\lambda\lambda$4686 and \ion{He}{I} $\lambda$4922 difficult. The emission component of the \ion{He}{I}$\lambda$4922 line that emerged at later times was weak at early times.

It is therefore challenging to determine the presence of He in SNe~Icn with only low-S/N optical spectra obtained near peak light, which may be why \citet{Pellegrino2022} did not clearly detect He in their sample\footnote{\citet{Pellegrino2022} claimed a \ion{He}{II} detection for SN~2019jc, but the feature is similar in strength to several nearby unidentified lines.}.  Our detections of He in SN~2022ann were either in NIR spectra (using the strongest He line, which is in the NIR at $\lambda$1.083~$\mu$m) or at later times in optical spectra.  \citet{Fraser2021} and \citet{Perley2022} both detected He in SN~2021csp in NIR spectra and high-S/N early optical spectra, respectively.  Robust detection of He in SNe~Icn may require high-S/N optical spectra near peak, NIR spectra, or late-time optical spectra.

Clear He detections in SNe~2021csp and 2022ann, as well as a potential detection in SN~2019jc, directly indicate that He is present in their circumstellar environments. Modeling of late-time spectra of SN~2019hgp did not support the presence of \ion{He}{I}; however, \citet{Gal-Yam2022} were unable to rule out a contribution from \ion{He}{II} in earlier spectra. SN~2021ckj lacks the data necessary for a clean detection. With this small sample, we conclude that a large fraction of, and perhaps all, SNe~Icn contain He in their circumstellar environments. Early-time, high-S/N, and NIR observations of future SNe~Icn will be critical for determining the ubiquity of He in this class.

\subsection{Heterogeneity in SNe~Icn}\label{sec:heterogeneity}
Although the sample is still small, SNe~Icn already appear to be highly diverse in several of their properties.  For instance, the current sample of five SNe~Icn have a range of luminosities at peak that is nearly the same as that for the much larger SN~Ibn sample \citep{Hosseinzadeh2017}.  The luminosity differences persist at late times ($t > 30$~days).  The early photometric evolution also appears to be quite heterogeneous with SN~2019hgp having a smooth rise and fall, SN~2021csp having a fast rise and linear decline, and SN~2022ann having nearly a plateau.  However, at late times, SN~Icn decline rates are more similar.

Despite the photometric diversity, the SN~Icn spectra are remarkably similar at maximum brightness; all SNe~Icn have blue continua, narrow P-Cygni features from C and O, and a lack of H and strong He lines or other features. SN~2021csp displays higher-ionization lines than the other SNe~Icn, but the generic features persist. All SNe~Icn with maximum-light spectra exhibit strong, narrow P-Cygni lines regardless of photometric properties.  In contrast to SNe~Ibn, no SN~Icn has yet been discovered that has only emission features (lacking a P-Cygni absorption component). These similarities suggest similar compositions and dynamics in SN~Icn circumstellar environments.

On the other hand, the spectral behaviour of SNe~Icn diverges after peak light. All SNe~Icn with late-time spectra, apart from SN~2022ann, eventually developed broad features as some, if not all, of the narrow interaction lines disappeared.  Meanwhile, we never see broad lines from SN~2022ann, indicating either a very low explosion velocity that is similar to that of the pre-explosion wind ($\sim$800~km~s$^{-1}$) or that the SN ejecta are veiled throughout our observations.  In all of our spectra of SN~2022ann, there are signs of continued circumstellar interaction with a C/O-rich CSM. SNe~Icn also show differing amounts of Fe fluorescence.

A possible explanation for the diverse late-time behaviour is a shared but asymmetric configuration of CSM causing a viewing-angle dependence on the optical-depth evolution. This scenario is able to explain observed spectroscopic diversity, while maintaining similar global CSM properties across SNe~Icn. For example, a toroidal configuration of CSM is commonly invoked to explain spectroscopic diversity observed in SNe~Ibn, and could explain the differing times at which the SN~Icn ejecta become visible. However, a torus of C- and O-rich CSM observed face-on would produce broad emission features in early-time spectra, a phenomenon not observed in SNe~Icn so far.

Another possibility is that the circumstellar environments around SNe~Icn, and potentially the mechanisms by which they are created, are themselves heterogeneous. The amount of Fe fluorescence could be tied to the host-galaxy metallicity, and therefore the abundance of Fe in the CSM. Assuming a mass-metallicity relationship for the host galaxies, the SNe~Icn shown in Figure \ref{fig:samplebreakdown} exhibit increasing levels of Fe fluorescence with higher-metallicity hosts. However, Fe fluorescence can be affected by a number of other factors besides abundance, such as CSM density and geometry. While simple modeling of SN~Icn bolometric light curves has yielded relatively similar explosion properties throughout the subclass \citep{Pellegrino2022}, more rigorous theoretical modeling of these systems could reveal global differences in their circumstellar environments or point to different progenitor scenarios entirely. More late-time observations and a larger sample of SNe~Icn could also help better constrain the relative contributions of radioactive decay and CSI to the light curves.

\subsection{Progenitor System}\label{sec:progenitor}
The observables of SN~2022ann provide constraints on its progenitor system. The low absorption velocities, total mass of the system, and host-galaxy environment all independently indicate that a massive WR star cannot be the progenitor of SN~2022ann. Therefore, we favour a binary progenitor scenario.

The strong, narrow C and O lines and lack of H and strong He lines in the spectra of SN~2022ann indicate that it is the result of an explosion of a star in a C/O-rich and H/He-poor circumstellar environment, consistent with what others have concluded for SNe~Icn \citep{Gal-Yam2022, Fraser2021, Perley2022, Pellegrino2022}.  Although we do not detect broad lines consistent with being SN ejecta, it is probable that the ejecta are also H/He poor, similar to other SNe~Icn.  These two properties shape our basic view of the progenitor system: a highly stripped star exploding within a dense, H/He-poor CSM.

SN~2022ann has two systemic velocities, $\sim$800~km~s$^{-1}$ from the P-Cygni absorption and $\sim$4900~km~s$^{-1}$ from the expansion of the photospheric radius derived from the bolometric light curve.  Although these could be different measurements of a highly asymmetric system, it is more likely that the photospheric velocity comes from a physically different component than the absorption velocity.  This scenario is possible if low-density CSM is ejected at the lower velocity well before the SN and the high-velocity SN ejecta is beneath the CSM outflow.

In the case of a constant stellar wind, we can place a limit on the extent of the CSM.  P-Cygni features are present until at least 34~days after explosion.  Assuming an SN velocity of 4900~km~s$^{-1}$, the CSM must extend to at least $r \approx 10^{15}$~cm.  If the CSM was a single ejection event and represents a thin shell with an expansion velocity of 800~km~s$^{-1}$, the ejection must have occurred $>$200~days before explosion.  If we assume a shock velocity of 30,000~km~s$^{-1}$, the shock would catch up to a CSM at $10^{15}$~cm in 5~days, exactly coincident with our first spectrum.  Therefore, it is unfortunately impossible to distinguish between a wind and an outburst scenario.

Assuming the measured absorption velocity ($\sim$ 800 km s$^{-1}$)is the escape velocity of a radiatively driven wind, we measure
\begin{equation}
  \frac{M}{R} = 1.67 \frac{{\rm M}_{\sun}}{{\rm R}_{\sun}}\, .
\end{equation}
The velocity and the resulting mass-to-radius ratio is much smaller than that of a Wolf-Rayet star \citep{Sander19}.  Additionally, the two Galactic WO2 stars, WR~102 and WR~142, have $M/R > 30$, but the WO3 WR~93b has $M/R \approx 18$ (with all having terminal velocities of $v_{\infty} \approx 5000$~km~s$^{-1}$).  Low-mass (1~M$_{\sun}$) WO model stars have $M/R \approx 5$ \citep{Langer89}.  In contrast, the CSM for SN~2022ann has a wind velocity an order of magnitude lower, resulting in the low mass-to-radius ratio.  Such a low wind velocity requires either (1) an extended photosphere out of hydrostatic equilibrium, (2) an outflow that is not driven by radiation pressure, (3) a very asymmetric outflow where the line-of-sight velocity is a small fraction of the true velocity, or (4) the wind originating from a single, but non-WR star.  The last two options seem highly unlikely or completely unphysical.

At the end of their lives, core-collapse progenitor stars are particularly active \citep[e.g.,][]{Kilpatrick21, Jacobson-Galan22}.  SNe~Ibn appear to be extraordinarily active, with the progenitor stars of SN~2006jc and SN~2019uo having luminous outbursts only 1--2~yr before explosion \citep{Foley07, Pastorello07, Strotjohann21}.  Such an outburst would certainly cause a departure from hydrostatic equilibrium for at least some time, but then we would expect the velocity to be even higher than the escape velocity.  This scenario would require there to be additional high-velocity material at even larger radii that could be detected with late-time monitoring \citep{Tinyanont2016, Margutti2017, Mauerhan2018}. 

If we assume that the wind is radiatively driven at the Eddington luminosity, then we find
\begin{equation}
  T = {\rm 50,\!400}~{\rm K} \, \left ( \frac{\kappa}{0.2 {\rm ~cm}^{2}{\rm ~g}^{-1}} \right )^{-1/4} \left ( \frac{v}{800 {\rm ~km~s}^{-1}} \right ) \left ( \frac{M}{10~{\rm M}_{\sun}} \right )^{-1/4}.
\end{equation}
While the opacity may be significantly higher than that of electron scattering, this places a reasonable limit on the effective temperature of the progenitor in this scenario.  Although the star is more extended than a normal WR star, it is still very hot with the peak of its emission in the UV.  Nevertheless, ``cool'' WO stars may be the progenitors of some SNe~Icn.

Alternatively, the velocity seen may not be the escape velocity from the progenitor star, but the escape velocity from a binary system \citep[e.g., see][]{Tauris15}. Studies have shown that binary interaction can strip stars of their H/He-envelopes and drive runaway mass loss \citep{Podsialowski92, Yoon2010}. Modeling of ultrastripped SNe (USSNe) yield small ejecta and $^{56}$Ni masses comparable to those of SNe~Icn, which is inconsistent with WR progenitors \citep{Drout2013, De18, Yao20}. Detailed analysis of these objects supports binary progenitor systems. Binary systems have several evolutionary pathways along which mass transfer can happen. Therefore, heterogeneity in the resulting CSM and ultimately SNe~Icn would be expected.

CCSN progenitors typically have short lifetimes (up to a few tens of million years), where the SNe often occur during the star-formation event that formed the SN progenitor (where the SN can quench local star formation).  Among dwarf galaxies ($<10^8$~M$\odot$), the host galaxy of SN~2022ann has a low sSFR, which suggests that SN~2022ann outlived its epoch of star formation {\it or} exploded before other high-mass stars formed.  The latter is statistically unlikely and a larger sample of SNe~Icn with low-mass host galaxies will be able to test this possibility.  The former is intriguing since the lifetime of a typical starburst in a dwarf galaxy is $\sim$100~Myr \citep{Lee2009}, which would disfavour OB-type progenitors and favour a lower-mass progenitor star that might require a companion to fully strip its H/He envelope. 

The low host-galaxy mass and metallicity ($Z \approx 0.1$~Z$_{\odot}$) also suggests a low progenitor metallicity. Given its likely low metallicity and extreme stripping, the progenitor of SN~2022ann is less likely to have significant line-driven mass loss.  The alternatives are either mass loss to a companion or episodic, perhaps explosive mass loss.  The former is responsible for the mass loss of the majority of SN~Ibc~and~IIb progenitor stars \citep[e.g.,][]{Smith11, Yoon17}, but has trouble explaining the degree of mass stripping necessary for SN~2022ann.  The latter has been seen in two SNe~Ibn \citep{Foley07, Pastorello07, Strotjohann21} and several SNe~IIn \citep{Smith2010, Foley2011, Mauerhan2013, Ofek2014, Strotjohann21}, and inferred for several other SNe with dense CSM \citep[see][for a review]{Smith2014}.  It is natural to expect a similar mass-loss mechanism for SN~2022ann; however, the absorption velocities are lower than expected for this case.  A combination of nonconservative mass transfer to a binary star, which could create the low-velocity outflow, and episodic mass loss, which could cause the extreme stripping, may be necessary for SN~2022ann.

\section{Conclusions}\label{sec:conc}

We have presented optical photometry and optical/NIR spectroscopy of SN~2022ann, the fifth reported SN~Icn, and its host galaxy, SDSS~J101729.72--022535.6. SN~2022ann has several unique and extreme properties relative to the other members of this small subclass.  Our observations of SN~2022ann provide unique insight into the origins of the rarest SN explosions, and undiscovered endpoints of stellar evolution.

The denominative characteristics of SNe~Icn, including SN~2022ann, are early-time optical spectra with blue continua that show narrow P-Cygni lines of C and O while lacking (strong) H and He features. The absorption minimum velocities of these features measure at $~\sim$1000~km~s$^{-1}$ across the SNe~Icn.

SN~2022ann has a uniquely constant brightness at early times and a relatively rapid late-time decline in redder bands after this plateau.  At peak it is also has relatively low luminosity compared to other known SNe~Icn, which have diverse peak luminosities as a whole. We model the bolometric light curve of SN~2022ann, finding that it is well described by 1.73~M$_{\odot}$ of ejecta interacting with 0.19~M$_{\odot}$ of CSM. We place a conservative upper limit on the $^{56}$Ni mass at 0.04~M$_{\odot}$.

Through its photospheric phase, the spectra of SN~2022ann are well described by a blackbody continuum with narrow P-Cygni lines primarily from C and O.  Unlike other SNe~Ibn and Icn, SN~2022ann never shows broad lines that could be associated with SN ejecta, nor a high blue continuum flux from Fe fluorescence. The lack of any broad lines despite a high velocity for the photosphere is perplexing.  The lack of Fe fluorescence suggests either (1) the Fe abundance is low, or (2) there is an insufficient flux of high-energy photons necessary to pump the electrons.

While we do not clearly detect any He emission in our early-time optical spectra of SN~2022ann, we detect He lines in both of our NIR spectra (first epoch at +10.1~days) and our later optical spectra (starting at 28.6~days).  Other SNe~Icn have weak He lines in their early spectra \citep{Fraser2021, Perley2022, Pellegrino2022}.  While some SNe~Icn have no clear detection of He, those cases lack high-S/N early spectra NIR spectra, and late-time optical spectra.  We posit that He may be ubiquitous in SNe~Icn but specific observations are necessary to detect it. A focus on obtaining these observations for future SNe~Icn will be necessary to measure the amount of He in these systems.

Whereas other SNe~Icn have CSM velocities consistent with WR winds ($\sim$1000--2000 km~s$^{-1}$) \citep{Gal-Yam2022, Fraser2021, Perley2022, Pellegrino2022}, the CSM of SN~2022ann has a velocity of only $\sim$800~km~s$^{-1}$, inconsistent with known WO stars and below the escape velocity for a compact massive star that is necessary to avoid strong H emission.  The progenitor star of SN~2022ann may have been ``puffed up'' by an outburst and out of hydrostatic equilibrium before explosion.  Alternatively, the wind velocity may be indicative of the escape velocity from a binary system rather than from the progenitor star itself.

The host galaxy of SN~2022ann is a low-mass dwarf galaxy with $\log(M/{\rm M}_\odot) = 7.34^{+0.11}_{-0.30}$ and a low SFR of $\log({\rm SFR}) = -2.20^{+0.30}_{-0.09}$~M$_{\odot}$~yr$^{-1}$. The dwarf-galaxy nature of SN~2022ann is likely linked to this extreme environment. In particular, the galaxy-averaged metallicity of $\sim$0.1~Z$_\odot$ suggests that the progenitor of SN~2022ann likely had low metallicity, making line-driven mass loss inefficient and unlikely to fully strip the H/He envelope from the progenitor star.  A binary companion could provide a mechanism for the necessary enhanced mass loss.  

Given the low CSM velocity, low $^{56}$Ni and ejecta masses, and SFR-poor host-galaxy environment, we favour a binary-stripping progenitor scenario for SN~2022ann over a single massive WR progenitor. The rarity of SNe~Icn may indicate that they are created during a brief or uncommon stage in binary evolution. Observations of future SNe~Icn, particularly at late times and in the NIR, will be critical for constraining the nature of this path of stellar evolution.

\section*{Acknowledgements}
Research at UC Santa Cruz is conducted on the unceded territory of Awaswas-speaking Uypi Tribe. The Amah Mutsun Tribal Band, comprised of the descendants of indigenous people taken to missions Santa Cruz and San Juan Bautista during Spanish colonisation of the Central Coast, is today working hard to restore traditional stewardship practices on these lands and heal from historical trauma. Keck I/II, ATLAS, and PS1 observations were conducted on the stolen land of the k\={a}naka `\={o}iwi people. We stand in solidarity with the Pu'uhonua o Pu'uhuluhulu Maunakea in their effort to preserve these sacred spaces for native Hawai`ians. Shane 3~m observations at Lick Observatory were conducted on the stolen land of the Ohlone (Costanoans), Tamyen, and Muwekma Ohlone tribes. Hobbly-Eberly Telescope observations were conducted on the stolen land of the Chiricahua Apache Nation,the Chiricahua Apache and Lipan Apache tribes, and Jumanos people. MMT observations were conducted on the stolen land of the Tohono O’odham and Hia-Ced O’odham nations, the Ak-Chin Indian Community, and Hohokam people.

The UCSC team is supported in part by NASA grant 80NSSC20K0953, NSF grant AST--1815935, the Gordon \& Betty Moore Foundation, the Heising-Simons Foundation, and by a fellowship from the David and Lucile Packard Foundation to R.J.F.
C.R.A.\ was supported by a VILLUM FONDEN Young Investigator Grant (project number 25501).  The DARK team was supported in part by a VILLUM FONDEN Investigator grant (project number 16599).  D.A.C.\ acknowledges support from the National Science Foundation (NSF) Graduate Research Fellowship under grant DGE--1339067.  D.O.J.\ is supported by NASA through Hubble Fellowship grant HF2-51462.001 awarded by the Space Telescope Science Institute (STScI), which is operated by the Association of Universities for Research in Astronomy, Inc., for NASA, under contract NAS5--26555. P.L.K.\ acknowledges support through NSF grant AST-1908823.  The Las Cumbres group is supported by AST--1911151 and AST--1911225.  M.R.S.\ is supported by the STScI Postdoctoral Fellowship. C.R.B.\ acknowledges the financial support from CNPq (316072/2021-4) and from FAPERJ (grants 201.456/2022 and 210.330/2022).  Support for A.V.F.'s group at U.C. Berkeley is provided by the Christopher R. Redlich Fund and many individual donors.  L.G.\ acknowledges financial support from the Spanish Ministerio de Ciencia e Innovaci\'on (MCIN), the Agencia Estatal de Investigaci\'on (AEI) 10.13039/501100011033, and the European Social Fund (ESF) ``Investing in your future'' under the 2019 Ram\'on y Cajal program RYC2019-027683-I and the PID2020-115253GA-I00 HOSTFLOWS project, from Centro Superior de Investigaciones Cient\'ificas (CSIC) under the PIE project 20215AT016, and the program Unidad de Excelencia Mar\'ia de Maeztu CEX2020-001058-M.  F.R.H.\ acknowledges funding from FAPESP grants 2018/21661-9 and 2021/11345-5.   C.G.\ is supported by a VILLUM FONDEN Young Investigator Grant (project number 25501).  C.L.\ is supported by the NSF Graduate Research Fellowship under grant DGE--2233066.

The Young Supernova Experiment (YSE) and its research infrastructure is supported by the European Research Council under the European Union's Horizon 2020 research and innovation programme (ERC Grant Agreement 101002652, PI K.\ Mandel), the Heising-Simons Foundation (2018-0913, PI R.\ Foley; 2018-0911, PI R.\ Margutti), NASA (NNG17PX03C, PI R.\ Foley), NSF (AST-1720756, AST-1815935, PI R.\ Foley; AST-1909796, AST-1944985, PI R.\ Margutti), the David \& Lucille Packard Foundation (PI R.\ Foley), VILLUM FONDEN (project 16599, PI J.\ Hjorth), and the Center for AstroPhysical Surveys (CAPS) at the National Center for Supercomputing Applications (NCSA) and the University of Illinois Urbana-Champaign.

A subset of the data presented herein were obtained at the W.\ M.\ Keck Observatory, which is operated as a scientific partnership among the California Institute of Technology, the University of California, and the NASA. The Observatory was made possible by the generous financial support of the W.\ M.\ Keck Foundation. The authors wish to recognise and acknowledge the very significant cultural role and reverence that the summit of Maunakea has always had within the indigenous Hawaiian community.  We are most fortunate to have the opportunity to conduct observations from this mountain.
A major upgrade of the Kast spectrograph on the Shane 3~m telescope at Lick Observatory was made possible through generous gifts from the Heising-Simons Foundation as well as William and Marina Kast. Research at Lick Observatory is partially supported by a generous gift from Google. 
 
Pan-STARRS is a project of the Institute for Astronomy of the University of Hawaii, and is supported by the NASA SSO Near Earth Observation Program under grants 80NSSC18K0971, NNX14AM74G, NNX12AR65G, NNX13AQ47G, NNX08AR22G, 80NSSC21K1572 and by the State of Hawaii.  The Pan-STARRS1 Surveys (PS1) and the PS1 public science archive have been made possible through contributions by the Institute for Astronomy, the University of Hawaii, the Pan-STARRS Project Office, the Max-Planck Society and its participating institutes, the Max Planck Institute for Astronomy, Heidelberg and the Max Planck Institute for Extraterrestrial Physics, Garching, The Johns Hopkins University, Durham University, the University of Edinburgh, the Queen's University Belfast, the Harvard-Smithsonian Center for Astrophysics, the Las Cumbres Observatory Global Telescope Network Incorporated, the National Central University of Taiwan, STScI, NASA under grant NNX08AR22G issued through the Planetary Science Division of the NASA Science Mission Directorate, NSF grant AST-1238877, the University of Maryland, Eotvos Lorand University (ELTE), the Los Alamos National Laboratory, and the Gordon and Betty Moore Foundation.

Based in part on observations obtained at the Southern Astrophysical Research (SOAR) telescope (program 2021A-0242 (SOAR2022A-001); PI Dimitriadis), which is a joint project of the Minist\'{e}rio da Ci\^{e}ncia, Tecnologia e Inova\c{c}\~{o}es (MCTI/LNA) do Brasil, the US NSF’s NOIRLab, the University of North Carolina at Chapel Hill (UNC), and Michigan State University (MSU).
Based in part on observations obtained at the international Gemini Observatory (program LP-204; PI Jacobson-Galan), a program of NSF’s NOIRLab, which is managed by the Association of Universities for Research in Astronomy (AURA) under a cooperative agreement with the NSF on behalf of the Gemini Observatory partnership: the NSF (United States), National Research Council (Canada), Agencia Nacional de Investigaci\'{o}n y Desarrollo (Chile), Ministerio de Ciencia, Tecnolog\'{i}a e Innovaci\'{o}n (Argentina), Minist\'{e}rio da Ci\^{e}ncia, Tecnologia, Inova\c{c}\~{o}es e Comunica\c{c}\~{o}es (Brazil), and Korea Astronomy and Space Science Institute (Republic of Korea).

This work makes used of data from the Las Cumbres Observatory network. This publication has made use of data collected at Lulin Observatory, partly supported by MoST grant 108-2112-M-008-001.  The S-PLUS project, including the T80-South robotic telescope and the S-PLUS scientific survey, was founded as a partnership between the Funda\c{c}\~{a}o de Amparo \`{a} Pesquisa do Estado de S\~{a}o Paulo (FAPESP), the Observat\'{o}rio Nacional (ON), the Federal University of Sergipe (UFS), and the Federal University of Santa Catarina (UFSC), with important financial and practical contributions from other collaborating institutes in Brazil, Chile (Universidad de La Serena), and Spain (Centro de Estudios de F\'{\i}sica del Cosmos de Arag\'{o}n, CEFCA). We further acknowledge financial support from the São Paulo Research Foundation (FAPESP), the Brazilian National Research Council (CNPq), the Coordination for the Improvement of Higher Education Personnel (CAPES), the Carlos Chagas Filho Rio de Janeiro State Research Foundation (FAPERJ) and the Brazilian Innovation Agency (FINEP). 

This work has made use of data from the Asteroid Terrestrial-impact Last Alert System (ATLAS) project. The Asteroid Terrestrial-impact Last Alert System (ATLAS) project is primarily funded to search for near-Earth objects (NEOs) through NASA grants NN12AR55G, 80NSSC18K0284, and 80NSSC18K1575; byproducts of the NEO search include images and catalogs from the survey area. This work was partially funded by Kepler/K2 grant J1944/80NSSC19K0112 and HST GO-15889, and STFC grants ST/T000198/1 and ST/S006109/1. The ATLAS science products have been made possible through the contributions of the University of Hawaii's Institute for Astronomy, the Queen’s University Belfast, the Space Telescope Science Institute, the South African Astronomical Observatory, and The Millennium Institute of Astrophysics (MAS), Chile.

The data presented here were obtained [in part] with ALFOSC, which is provided by the Instituto de Astrofisica de Andalucia (IAA) under a joint agreement with the University of Copenhagen and NOT.

Observations reported here were obtained in part at the MMT Observatory, a joint facility of the University of Arizona and the Smithsonian Institution.
The Hobby-Eberly Telescope (HET) is a joint project of the University of Texas at Austin, the Pennsylvania State University, Ludwig-Maximilians-Universit\"at M\"unchen, and Georg-August-Universit\"at G\"ottingen. The HET is named in honor of its principal benefactors, William P.\ Hobby and Robert E.\ Eberly.

The Legacy Surveys consist of three individual and complementary projects: the Dark Energy Camera Legacy Survey (DECaLS; Proposal ID 2014B-0404; PIs David Schlegel and Arjun Dey), the Beijing-Arizona Sky Survey (BASS; Proposal ID 2015A-0801; PIs Zhou Xu and Xiaohui Fan), and the Mayall z-band Legacy Survey (MzLS; Proposal 2016A-0453; PI Arjun Dey). DECaLS, BASS, and MzLS together include data obtained (respectively) at the Blanco telescope, Cerro Tololo Inter-American Observatory, NSF’s NOIRLab; the Bok telescope, Steward Observatory, University of Arizona; and the Mayall telescope, Kitt Peak National Observatory, NOIRLab. Pipeline processing and analyses of the data were supported by NOIRLab and the Lawrence Berkeley National Laboratory (LBNL). The Legacy Surveys project is honored to be permitted to conduct astronomical research on Iolkam Du’ag (Kitt Peak), a mountain with particular significance to the Tohono O’odham Nation.
NOIRLab is operated by the Association of Universities for Research in Astronomy (AURA) under a cooperative agreement with the NSF. LBNL is managed by the Regents of the University of California under contract to the U.S.\ Department of Energy.

This project used data obtained with the Dark Energy Camera (DECam), which was constructed by the Dark Energy Survey (DES) collaboration. Funding for the DES Projects has been provided by the U.S. Department of Energy, the U.S. NSF, the Ministry of Science and Education of Spain, the Science and Technology Facilities Council of the United Kingdom, the Higher Education Funding Council for England, the National Center for Supercomputing Applications at the University of Illinois at Urbana-Champaign, the Kavli Institute of Cosmological Physics at the University of Chicago, Center for Cosmology and Astro-Particle Physics at the Ohio State University, the Mitchell Institute for Fundamental Physics and Astronomy at Texas A\&M University, Financiadora de Estudos e Projetos, Fundacao Carlos Chagas Filho de Amparo, Financiadora de Estudos e Projetos, Fundacao Carlos Chagas Filho de Amparo a Pesquisa do Estado do Rio de Janeiro, Conselho Nacional de Desenvolvimento Cientifico e Tecnologico and the Ministerio da Ciencia, Tecnologia e Inovacao, the Deutsche Forschungsgemeinschaft and the Collaborating Institutions in the Dark Energy Survey. The Collaborating Institutions are Argonne National Laboratory, the University of California at Santa Cruz, the University of Cambridge, Centro de Investigaciones Energeticas, Medioambientales y Tecnologicas-Madrid, the University of Chicago, University College London, the DES-Brazil Consortium, the University of Edinburgh, the Eidgenossische Technische Hochschule (ETH) Zurich, Fermi National Accelerator Laboratory, the University of Illinois at Urbana-Champaign, the Institut de Ciencies de l’Espai (IEEC/CSIC), the Institut de Fisica d’Altes Energies, Lawrence Berkeley National Laboratory, the Ludwig Maximilians Universitat Munchen and the associated Excellence Cluster Universe, the University of Michigan, NSF’s NOIRLab, the University of Nottingham, the Ohio State University, the University of Pennsylvania, the University of Portsmouth, SLAC National Accelerator Laboratory, Stanford University, the University of Sussex, and Texas A\&M University.

IRAF is distributed by NOAO, which is operated by AURA, Inc., under cooperative agreement with the NSF.

\bibliographystyle{mnras}
\bibliography{ref}

\newpage 

\appendix

\section{Appendix}

\begin{table}
    \centering
    \caption{Log of photometric observations of SN~2022ann. $UBV$ filters reported in Vega-based magnitudes, while all other filters are reported in AB magnitudes. Data provided in table are not extinction-corrected. Magnitude uncertainties are presented in parentheses following the magnitudes. A portion of the data is shown here; the full data will be provided via an electronically-readable table online.} 
     \begin{tabular}{cccc}
\hline
\hline
MJD & Instrument & Filter & Apparent Magnitude\\
\hline
59604.5 & ATLAS-ACAM1 & o & 19.22 (0.11)\\
59605.6 & ATLAS-ACAM1 & o & 19.14 (0.11)\\
... & ... & ... & ... \\
\hline
     \end{tabular}\label{tab:phottable}
\end{table}

\begin{table*}
\centering
\caption{Log of spectroscopic observations for SN~2022ann.}
\label{tab:speclog}
\begin{tabular}{lccccc}
\hline
\hline
UT Date & Phase (days) & Telescope & Instrument &Range (\AA) & Exp. Time (s)\\
\hline
2022 February 06.27 & +2.8 & Shane & Kast & 3250--10,380 & 2400\\
2022 February 08.27 & +4.8 & SOAR & Goodman & 3810--6690 & 1800\\
2022 February 13.57 & +10.1 & Keck-II & NIRES & 9200--23,500 & 1200\\
2022 February 13.97 & +10.5 & NOT & ALFOSC & 3620--8575 & 3600\\
2022 February 20.17 & +16.7 & SOAR & Goodman & 4815--8600 & 2400\\
2022 February 22.57 & +19.1 & Keck-II & NIRES & 9200--23,500 & 2400\\
2022 February 23.97 & +20.5 & SOAR & Goodman & 3705--8565 & 1800\\
2022 February 25.97 & +22.5 & HET & LRS2 & 3480--6610 & 1200\\
2022 March 01.27 & +25.8 & Shane & Kast & 3130--9615 & 4800\\
2022 March 04.07 & +28.6 & Keck-I & LRIS & 2990--9790 & 1500\\
2022 March 07.07 & +31.6 & MMT & Binospec & 3670--8760 & 2000\\
2022 April 08.07 & +63.6 & Gemini-S & GMOS & 5270--8950 & 3000\\
2022 April 25.27 & +80.8 & Keck-I & LRIS & 3600--9670 & 3200\\
\hline
\end{tabular}
\end{table*}

\begin{table}
    \centering
    \caption{Host-galaxy photometry measured from pre-imaging all-sky public imaging surveys. Photometry is not corrected for Galactic foreground extinction. Upper limits are 2$\sigma$. \textbf{Notes.} [1] \citet{SDSS+2000}, [2] \citet{DESI+LIS+2019}, [3] \citet{Flewelling16}.} 
     \begin{tabular}{lccllc}
\hline
\hline
Filter  &   AB mag   &   Uncertainty  & Survey & Reference \\
\hline
$u'$   & $>$22.37 & -- &  SDSS DR16 & [1] \\

$g'$   & 22.77 & 0.12 &  DESI Legacy Imaging & [2] \\
$g'$   & 22.72 & 0.27 &  SDSS DR16 & [1] \\
$g'$   & 22.72 & 0.39 &  Pan-STARRS & [3] \\

$r'$   & 22.38 & 0.16 &  DESI Legacy Imaging & [2] \\
$r'$   & 22.21 & 0.25 &  SDSS DR16 & [1] \\
$r'$   & 22.40 & 0.37 &  Pan-STARRS & [3] \\

$i'$   & 22.10 & 0.27 &  SDSS DR16 & [1] \\
$i'$   & 22.25 & 0.36 &  Pan-STARRS 3PI & [3] \\

$z'$   & 21.97    & 0.16 &  DESI Legacy Imaging & [2] \\
$z'$   & $>$19.92 & -- &  SDSS DR16 & [1] \\
$z'$   & 21.92     & 0.31 &  Pan-STARRS & [3] \\

$y'$  &$>$21.51 & --  &  Pan-STARRS & [3] \\
\hline
     \end{tabular}\label{tab:hostphotometry}
\end{table}

\section{Author Affiliations}
\label{appendix:affil}
$^{\ucsc}$Department of Astronomy and Astrophysics, University of California, Santa Cruz, CA 95064, USA\\
$^{\psu}$Department of Astronomy \& Astrophysics, The Pennsylvania State University, University Park, PA 16802, USA\\
$^{\dark}$DARK, Niels Bohr Institute, University of Copenhagen, Jagtvej 128, 2200 Copenhagen, Denmark\\
$^{\uiuc}$Department of Astronomy, University of Illinois at Urbana-Champaign, 1002 W. Green St., IL 61801, USA\\
$^{\ifa}$Institute for Astronomy, University of Hawaii, 2680 Woodlawn Drive, Honolulu, HI 96822, USA\\
$^{\gemini}$Gemini Observatory/NSF's NOIRLab, 670 N. A'ohoku Place, Hilo, Hawai'i, 96720, USA\\
$^{\umn}$Minnesota Institute for Astrophysics, University of Minnesota, 116 Church Street SE, Minneapolis, MN 55455, USA\\
$^{\ciera}$Center for Interdisciplinary Exploration and Research in Astrophysics (CIERA) and Department of Physics and Astronomy, Northwestern University, \\Evanston, IL 60208, USA\\
$^{\ncu}$Graduate Institute of Astronomy, National Central University, 300 Zhongda Road, Zhongli, Taoyuan 32001, Taiwan\\
$^{\lco}$Las Cumbres Observatory, 6740 Cortona Drive, Suite 102, Goleta, CA 93117-5575, USA\\
$^{\ucsb}$Department of Physics, University of California, Santa Barbara, CA 93106-9530, USA\\
$^{\jhu}$Department of Physics and Astronomy, The Johns Hopkins University, Baltimore, MD 21218, USA\\
$^{\stsci}$Space Telescope Science Institute, Baltimore, MD 21218, USA\\
$^{\het}$Hobby Eberly Telescope, University of Texas, Austin, TX 78712, USA\\
$^{\mel}$School of Physics, The University of Melbourne, VIC 3010, Australia\\
$^{\arc}$ARC Centre of Excellence for All Sky Astrophysics in 3 Dimensions (ASTRO 3D)\\
$^{\cbpf}$Centro Brasileiro de Pesquisas F\'isicas, Rua Dr. Xavier Sigaud 150, CEP 22290-180, Rio de Janeiro, RJ, Brazil\\
$^{\rut}$Department of Physics and Astronomy, Rutgers, the State University of New Jersey, 136 Frelinghuysen Road, Piscataway, NJ 08854-8019, USA\\
$^{\ucb}$Department of Astronomy, University of California, Berkeley, CA 94720-3411, USA\\
$^{\ice}$Institute of Space Sciences (ICE, CSIC), Campus UAB, Carrer de Can Magrans, s/n, E-08193 Barcelona, Spain.\\
$^{\ieec}$Institut d’Estudis Espacials de Catalunya (IEEC), E-08034 Barcelona, Spain.\\
$^{\sp}$Department of Astronomy, University of Sao Paulo, Rua do Matao, 1226, 05509-900 Sao Paulo, Brazil\\
$^{\seti}$SETI Institute, 339 N. Bernardo Ave., Ste. 200, Mountain View, CA 94043, USA\\
$^{\ufrgs}$Departamento de Astronomia, Instituto de F\'isica, Universidade Federal do Rio Grande do Sul (UFRGS), Av. Bento Goncalves 9500.\\
$^{\noao}$NOAO, P.O. Box 26732, Tucson, AZ 85726, USA\\
$^{\qub}$Astrophysics Research Centre, School of Mathematics and Physics, Queen’s University Belfast, Belfast BT7 1NN, UK\\
\bsp	
\label{lastpage}
\end{document}